\documentstyle[astrobib,epsfig,times]{mn-ab}

\title[MUNICS -- I. Field selection, object extraction, and
photometry]{The Munich Near-Infrared Cluster Survey -- I. Field
  selection, object extraction, and photometry}

\author[N. Drory et al.]  {
  N.~Drory,$^1$\footnotemark[1]\footnotemark[2]
  G.~Feulner,$^1$\footnotemark[1] 
  R.~Bender,$^1$\footnotemark[1]
  C.~S.~Botzler,$^1$ 
  U.~Hopp,$^1$\footnotemark[1] 
  C.~Maraston,$^1$
  \newauthor 
  C.~Mendes de Oliveira,$^2$\footnotemark[1]
  J.~Snigula$^1$\footnotemark[1]\\
  $^1$Universit\"ats-Sternwarte M\"unchen, Scheinerstr. 1, 
  D-81679 M\"unchen, Germany\\
  $^2$Instituto Astron\^omico e Geof\'{\i}sico, Av Miguel St\'efano
  4200, 04301-904, S\~ao Paulo, Brazil}

\date{Accepted --; Received --; }

\pagerange{\pageref{firstpage}--\pageref{lastpage}} \pubyear{2001}

\begin{document}

\label{firstpage}

\maketitle

%
%
\begin{abstract}
  The Munich Near-IR Cluster Survey (MUNICS) is a wide-area,
  medium-deep, photometric survey selected in the $K'$ band. It covers
  an area of roughly one square degree in the $K'$ and $J$ near-IR
  pass-bands. The survey area consists of $16$ $6\arcmin \times
  6\arcmin$ fields targeted at QSOs with redshifts $0.5 < z < 2$ and
  $7$ $28\arcmin \times 13\arcmin$ stripes targeted at `random' high
  Galactic latitude fields. Ten of the QSO fields were additionally
  imaged in $R$ and $I$, and $0.6~\mathrm{deg}^2$ of the randomly
  selected fields were also imaged in the $V$, $R$, and $I$ bands. The
  resulting object catalogues were strictly selected in $K'$, having a
  limiting magnitude ($50$ per cent completeness) of $K' \sim
  19.5$~mag and $J \sim 21$~mag, sufficiently deep to detect passively
  evolving systems up to a redshift of $z \la 1.5$ and luminosity of
  $0.5 L^*$. The optical data reach a depth of roughly $R \sim
  23.5$~mag. The project's main scientific aims are the identification
  of galaxy clusters at redshifts around unity and the selection of a
  large sample of field early-type galaxies at $0 < z < 1.5$ for
  evolutionary studies.  In this paper -- the first in a series -- we
  describe the survey's concept, the selection of the survey fields,
  the near-IR and optical imaging and data reduction, object
  extraction, and the construction of photometric catalogues.
  Finally, we show the $J\!-\!K'$ vs. $K'$ colour--magnitude diagramme
  and the $R\!-\!J$ vs. $J\!-\!K'$, $V\!-\!I$ vs. $J\!-\!K'$, and
  $V\!-\!I$ vs. $V\!-\!R$ colour--colour diagrammes for MUNICS
  objects, together with stellar population-synthesis models for
  different star-formation histories, and conclude that the data set
  presented is suitable for extracting a catalogue of massive field
  galaxies in the redshift range $0.5 \la z \la 1.5$ for evolutionary
  studies and follow-up observations.
\end{abstract}

\begin{keywords}
  surveys -- infrared: galaxies -- galaxies: photometry -- 
  galaxies: evolution -- cosmology: observations
\end{keywords}

\footnotetext[1]{Visiting astronomer at the German-Spanish
  Astronomical Center, Calar Alto, operated by the Max-Planck-Institut
  f\"ur Astronomie, Heidelberg, jointly with the Spanish National
  Commission for Astronomy.}  
\footnotetext[2]{Visiting astronomer at
  the McDonald Observatory, Ft.\ Davis, Texas, operated by the
  University of Texas at Austin.}

%
%

\section{Introduction}
\label{s:introduction}

Directly observing the evolution of individual galaxies with time is,
unfortunately, not possible. Therefore we must rely on investigating
the statistical properties of the whole galaxy population as a
function of redshift, trying to draw conclusions from ensemble
properties on the evolution of typical members of these ensembles, and
thus facing difficulties, like, for example, discriminating between
luminosity evolution and number density evolution.

Much work has been invested in this field, resulting in a lot of
progress in the last decade which has seen many imaging and redshift
surveys being undertaken using different selection techniques in
wave-bands from the UV to the sub-mm. These surveys have a wide range
of scientific applications, from the detection of high-redshift galaxy
clusters to the study of the evolution of `normal' field galaxies.

The earlier optically and near-IR selected redshift and imaging
surveys, among others \citeN{BES88}, \citeN{LDSS90}, \citeN{LCG91},
\citeN{GPCM94}, and \citeN{CGHSHW94}, laid the path to the landmark
CFRS (Canada-France Redshift Survey; \citeNP{CFRS95}), an $I$-band
selected redshift survey mapping the evolution of the galaxy
population out to $z \sim 1$. Also many `pencil-beam' surveys have
been carried out, the most prominent being the Hubble Deep Field North
and South \cite{HDF96,HDF00} and their ground-based imaging and
spectroscopic follow-ups, allowing us a first glimpse at the galaxy
population at $2.5 \la z \la 4.5$. The inability to determine
redshifts spectroscopically for all objects where multi-band imaging
data is available (because of limited telescope resources, either in
observing time or in collecting area), caused photometric redshift
determination techniques to gain attention again
\cite{Baum62,Koo85,FLY99,Benitez00}. This, and the wide-field imagers
becoming available in the optical and also in the near-infrared
wavelength regime, made multi-band imaging surveys a very promising
option for further studies in galaxy evolution.

Selection in in a single pass-band introduces different (and sometimes
subtle) selection effects, a well-known fact which need not
necessarily be considered at the disadvantage of the resulting object
database, as long as the selection function is well-understood and
under control. These selection effects can be used deliberately for
probing different galaxy populations and different aspects of their
evolution. While selection in blue pass-bands is used to study star
forming sources, selection in the near-IR is predominantly sensitive
to the light of old stellar populations. Near-IR $k$-corrections are
small even at redshifts above unity and insensitive to the spectral
type of the observed objects \cite{CGHSHW94} and to short-lived bursts
of star formation, as has been pointed out by \citeN{KC98a}. Thus
near-IR selected surveys are thought to be much less biased with
respect to the mix of spectral types compared to optically selected
surveys. Furthermore, the uncertainties resulting from inhomogeneous
dust absorption are minimal in the near-IR. It has therefore been
concluded that near-IR selection is a feasible attempt at a selection
in stellar mass \cite{RR93,BE00}.

The Munich Near-Infrared Cluster Survey (MUNICS) is an attempt at
closing the gap between previously undertaken infrared-selected deep
pencil-beam surveys \cite{GCW93,MBRTF95,CGHSHW94,Detal95,HDF96,SIGM97}
and relatively shallow wide-area surveys \cite{MSE93,GPCM94,GSFC97},
simultaneously profiting from the advantages of near-infrared
selection.

MUNICS is a wide-area, medium-deep, photometric survey selected in the
$K'$ band. One part of the surveyed fields was centred on known
quasars, while the rest was randomly selected at high Galactic
latitudes. It covers an area of roughly one square degree in the $K'$
and $J$ bands with optical follow-up imaging in the $I$, $R$, and $V$
bands for a large fraction of the total surveyed area.

The resulting object catalogues are strictly selected in $K'$ with a
limiting magnitude of $K' \sim 19.5$~mag and $J \sim 21$~mag,
sufficiently deep to detect passively evolving systems up to a
redshift of $z \la 1.5$ and luminosity of $0.5 L^*$ (see
Fig.~\ref{f:models_kjk}). The optical data reach a depth of roughly $R
\sim 23.5$~mag.

This paper is laid out as follows. Section~\ref{s:concept} deals with
the survey's scientific aims and concept. The required sensitivity in
terms of limiting magnitudes, the selection of the survey fields as
well as the photometric system adopted are described. In
Section~\ref{s:observations} we give an overview of the observations
and discuss the reduction of the near-IR and optical imaging data.
Section~\ref{s:analysis} contains the data analysis. The methods for
object detection, photometry, and object classification are
discussed, and a first analysis of the survey's completeness as well
as number counts in five colours are presented.

%
%

\section{Survey concept and layout}
\label{s:concept}


\subsection{Scientific aims}
\label{s:aims}

\begin{figure}
  \centerline{\epsfig{figure=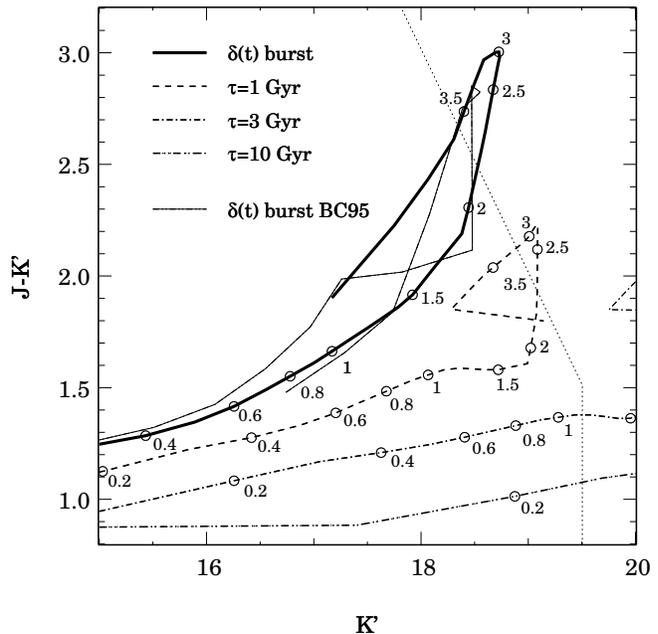,width=8.5cm}}
  \caption{Stellar population synthesis models in the $J\!-\!K'$ vs. $K'$ 
    plane for different star-formation histories, a ${\mathrm SFR}(t)
    \propto \delta(t)$ burst at $z=4$, and 3 exponential star
    formation rates ${\mathrm SFR}(t) \propto \exp(t/\tau)$ with $\tau
    = $ 1, 3, and 10 Gyr, setting in at $z=4$. The models are
    normalised to have a luminosity of $L^*_B$ at the present epoch,
    according to the type-dependent luminosity function of the Virgo
    cluster (see text) as given by Sandage, Binggeli, \& Tammann
    (1985).  The thin dotted line indicates the limiting depth (50 per cent 
    completeness) of the MUNICS data in $J$ and $K'$. 
    The adopted cosmology is $H_0=65,
    \Omega_0=0.3, \Omega_\Lambda=0.7$. The SSP models used in the
    synthesis are from Maraston (1998). Additionally, an SSP model by
    Bruzual \& Charlot (1995) is shown for comparison as a thin
    solid line.}
  \label{f:models_kjk}
\end{figure}
\nocite{SBT85,Maraston98,BC93}

The project's main scientific aims are the following. First, to
identify clusters of galaxies at high redshift by detecting their
luminous early-type galaxy population. As has been shown in the last
years, the early-type galaxy population in clusters is well in place
at redshifts of at least 0.8 \cite{SED95,SEESSD97,SED98,PSEDE99}.
Given the small $k$-corrections in the $K$ band, this makes selection
in the near-IR a promising approach to detect clusters at redshifts
around unity, complementing selection in other optical bands.
Clusters of galaxies allow to find large numbers of massive galaxies
at higher redshift and thus represent unique laboratories to study the
evolution of galaxies in high-density regions as a function of
redshift, and in contrast to the evolution of similar galaxies in the
field. Furthermore, the evolution of the number density of clusters is
a promising test of cosmological models, depending sensitively on the
density parameter $\Omega_0$ \cite{ECF96,BFC97,BF98,ECFH98}. While the
number of clusters known at redshifts $z > 0.5$ is steadily increasing
(mostly due to X-ray selection), samples selected {\em uniformly} in
the optical and near-IR wavelength ranges are still deficient.

Cluster detection at high redshifts is strongly biased towards the
most massive systems, mainly because of lack of detection sensitivity
for lower mass systems. Finding also less massive systems is important
when reasoning about hierarchical galaxy formation models, since the
galaxies in the densest environments formed earlier, so by looking
only at the most dense environments one is effectively pushing the
epoch of collapse, merging, and star formation out to higher redshifts
and further away from the observational window. Therefore we decided
to centre a subset of the MUNICS fields on known quasars hoping to
increase the chance of detecting clusters in their environment.

Secondly, a statistically well-defined sample of the early-type galaxy
population {\em in the field} can be constructed from our catalogues,
which will be used to study the evolutionary history of such objects
in the redshift range $0 < z < 1$ by means of the $K$-band selected
luminosity function, the luminosity density at near-infrared
wavelengths, and the two-point correlation function. Again, $K$-band
selection offers unique opportunities due to the close connection
between near-IR luminosity and stellar mass \cite{BE00}, and thus
allows direct assessment of the predictions of hierarchical galaxy
formation theories.

Thirdly, the nature of extremely red objects (EROs;
\citeNP{ERR88,HR94}) will be examined. EROs, usually defined in terms
of $R\!-\!K$ greater than or approximately equal to 5 at moderately
faint $K$-band magnitudes of $K \ge 18$, are thought to be either
high-redshift early-type galaxies or heavily extincted starburst
galaxies \cite{CDSPetal99,SIKetal99}, the relative contribution of the
two sub-populations being yet highly uncertain.  Due to the small
areas of the surveys available so far, even the surface density of
these objects is not reliably known \cite{Thompsonetal99}. Since they
mostly are $R$-band `dropouts', having the possibility to detect such
objects in the MUNICS data in the $I$ and $J$ bands, together with the
large field covered, will enable us to gain valuable information on
their nature.


\subsection{Limiting sensitivity}
\label{s:sensitivity}

Fig.~\ref{f:models_kjk} shows stellar population synthesis models in
the $J\!-\!K'$ vs. $K'$ plane for different star-formation histories,
a ${\mathrm SFR}(t) \propto \delta(t)$ Simple Stellar Population
(SSP), and 3 exponential star formation rates ${\mathrm SFR}(t)
\propto \exp(t/\tau)$ with $\tau = $ 1, 3, and 10 Gyr. The onset of
star formation occurs at $z=4$ in all models. The populations are
normalised to have a luminosity of $L^*_B$ at the present epoch,
according to the type-dependent luminosity function of the Virgo
cluster as given by \citeN{SBT85}. The adopted values for $M^*_B$ are
$M^*_B=-21.5$ for the SSP model (elliptical/S0 galaxy), $M^*_B=-20.5$
for $\tau=1$ (Sa--Sb spiral), $M^*_B=-19.5$ for $\tau=3$ (Sc), and
$M^*_B=-17.5$ for $\tau=10$ (Sd and later). The cosmology is
$H_0=65~{\mathrm km~s^{-1}~Mpc^{-1}}, \Omega_0=0.3,
\Omega_\Lambda=0.7$.

The SSP models are taken from \citeN{Maraston98}. The distinguishing
feature of that synthesis method is the adaption of the fuel
consumption theorem to evaluate the energetics of the post
main-sequence evolutionary phases.  The models used here have solar
metallicity and age ranging from 30 Myr to 15 Gyr. The Initial Mass
Function (IMF) is a power law $\Psi(M) \propto M^{-(1+x)}$ with
Salpeter exponent $x=1.35$ down to a lower mass limit of $0.1
M_{\sun}$. The optical and infrared colours predicted by these SSP
models are calibrated against Milky Way and Magellanic Cloud globular
clusters and compared to similar models from the literature in
\citeN{Maraston98}. Fig.~\ref{f:models_kjk} also shows an SSP model by
\citeN{BC93} using the 1995 version of their code, with solar
metallicity and a Salpeter IMF. The models evolve similarly up to
redshifts of $\sim 1$. The differences in colour are likely due to the
different treatment of the post main sequence stages and are discussed
in \citeN{Maraston98}.

Following the predictions of these models, the limiting magnitudes in
the near-IR wave-bands have been chosen to be 19.5~mag in $K'$ and
21.0~mag in $J$, such that early-type objects having luminosities of
$\ga 0.5 L^*$ at the present epoch can be detected in $K'$ virtually
at any redshift, and in $J$ up to a redshift of $z \la 1.5$, assuming
passive evolution.

This is in agreement with the findings of the CFRS, which has shown
that, while the luminosity function of the population of blue field
galaxies shows significant signs of evolution in the redshift range
$0.2 < z < 1$ -- explainable by brightening or increase in space
density -- the redder part of the population (roughly redder than Sbc)
shows no signs of evolution of its luminosity function in the same
redshift range \cite{CFRS695}. The latter is interpreted in terms of
brightening of the individual galaxies through passive evolution
counterbalanced by negative density evolution, such that the
luminosity function of the early-type population effectively does not
evolve.


\subsection{Field selection}
\label{s:fieldselection}

\begin{figure}
  \centerline{\epsfig{figure=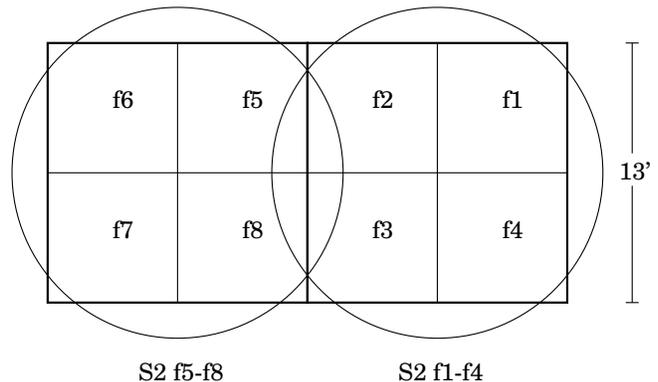,width=8.5cm}}
  \caption{Layout and nomenclature of one of the MUNICS `mosaic fields'. 
    The size of the stripe is $13\arcmin \times 28\arcmin$, covered by
    $2 \times 4$ pointings in the near-IR and two pointings in the
    optical with the circular field of view of CAFOS (see text). The
    IR image is divided for technical reasons into two $2 \times 2$
    mosaics. Each optical frame and IR mosaic frame are denoted by the
    name of the stripe, here S2, followed by the subfields they cover,
    giving S2 f1\dots f4 and S2 f5\dots f8 for this stripe.}
  \label{f:fgeom}
\end{figure}

The MUNICS survey consists of two sets of near-IR target fields, one
set of single camera pointings having an effective field of view of
$6\arcmin \times 6\arcmin$ pointed towards quasars, and a second set
of $28\arcmin \times 13\arcmin$ fields constructed from mosaics of
pointings targeted at random high Galactic latitude fields. This
second set of fields was selected to contain no bright stars, nearby
bright galaxies, and known nearby clusters of galaxies, and
furthermore, to have low Galactic reddening (which is all together
difficult to accomplish together with the prerequisite of having no
bright star within the field, given our field size).

A total of 16 fields targeted towards quasars with redshifts $0.5 < z
< 2$ were observed. These fields will be referred to as `quasar
fields' hereafter, labelled Q1\dots Q16. The quasars were selected
from the seventh edition of the \citeN{VCV96} catalogue.  The
selection criteria were $B < 19.0$~mag, $0\degr <$ Dec $< 65\degr$,
0$^{\mathrm h} <$ RA $< 18^{\mathrm h}$, and 0.5 $< z <$ 2. Six of
these quasars are not detected in the radio bands of the catalogue
(6~cm and 11~cm) and are therefore considered radio quiet. The
remaining 10 are radio loud.

A second set of 7 fields was targeted at high Galactic latitude
`empty' fields, i.e.\ free of bright stars ($V < 17$~mag) and known
nearby extragalactic objects. These fields will be called `mosaic
fields' hereafter, for they are mosaiced images in the near-IR. They
are labelled S1\dots S7.  Each such field is laid out as a stripe of
$4 \times 2$ IR pointings, yielding an area of $28\arcmin \times
13\arcmin$. For technical reasons, namely that four near-IR pointings
can be completed in $K'$ and $J$ during a single night as well as
image size and efficiency of optical follow-up observations (see
below), each such stripe is divided into two $2 \times 2$ mosaics of
single IR frames, denoted f1--f4 and f5--f8.  This particular geometry
was chosen for efficiency since such a $2 \times 2$ mosaic of IR
frames suits the circular field of view of the optical imager we used
(CAFOS, having a diameter of roughly 16~arcmin; see
Section~\ref{s:optical}). For clarity, Fig.~\ref{f:fgeom} shows a
sketch of the geometry and nomenclature of the mosaic fields.

Finally, Table~\ref{t:fields} lists the coordinates, available
pass-bands, seeing, and Galactic foreground extinction of the observed
mosaic and quasar fields.


\subsection{Photometric system}
\label{s:photsys}

\begin{figure}
  \centerline{\epsfig{figure=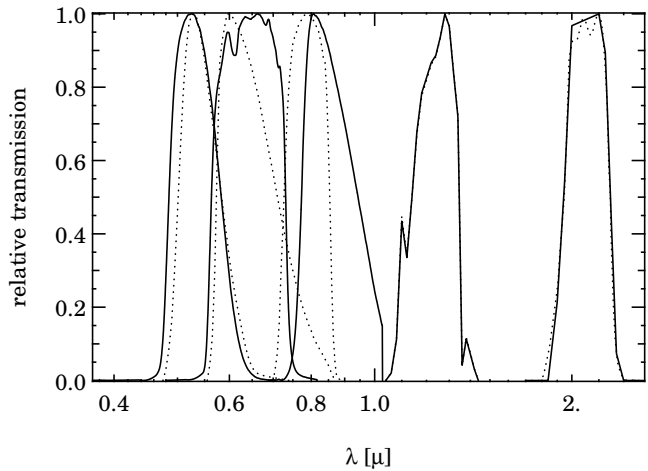,width=8.5cm}}
  \caption{Relative transmission of MUNICS $V,R,I,J$, and $K'$ 
    filter (solid curves) curves including quantum efficiency of the
    CCD ($V$,$R$, and $I$) and Rockwell HAWAII near-IR array ($J$ and
    $K'$), as well as the atmospheric transmission in the near-IR.
    Relative transmission curves of standard Johnson-Kron-Cousins $V$,
    $R$, and $I$ filters, as well as $J$ and $K'$ are shown for
    comparison (dotted curves).}
  \label{f:mun_filters}
\end{figure}

\begin{figure}
    \epsfig{figure=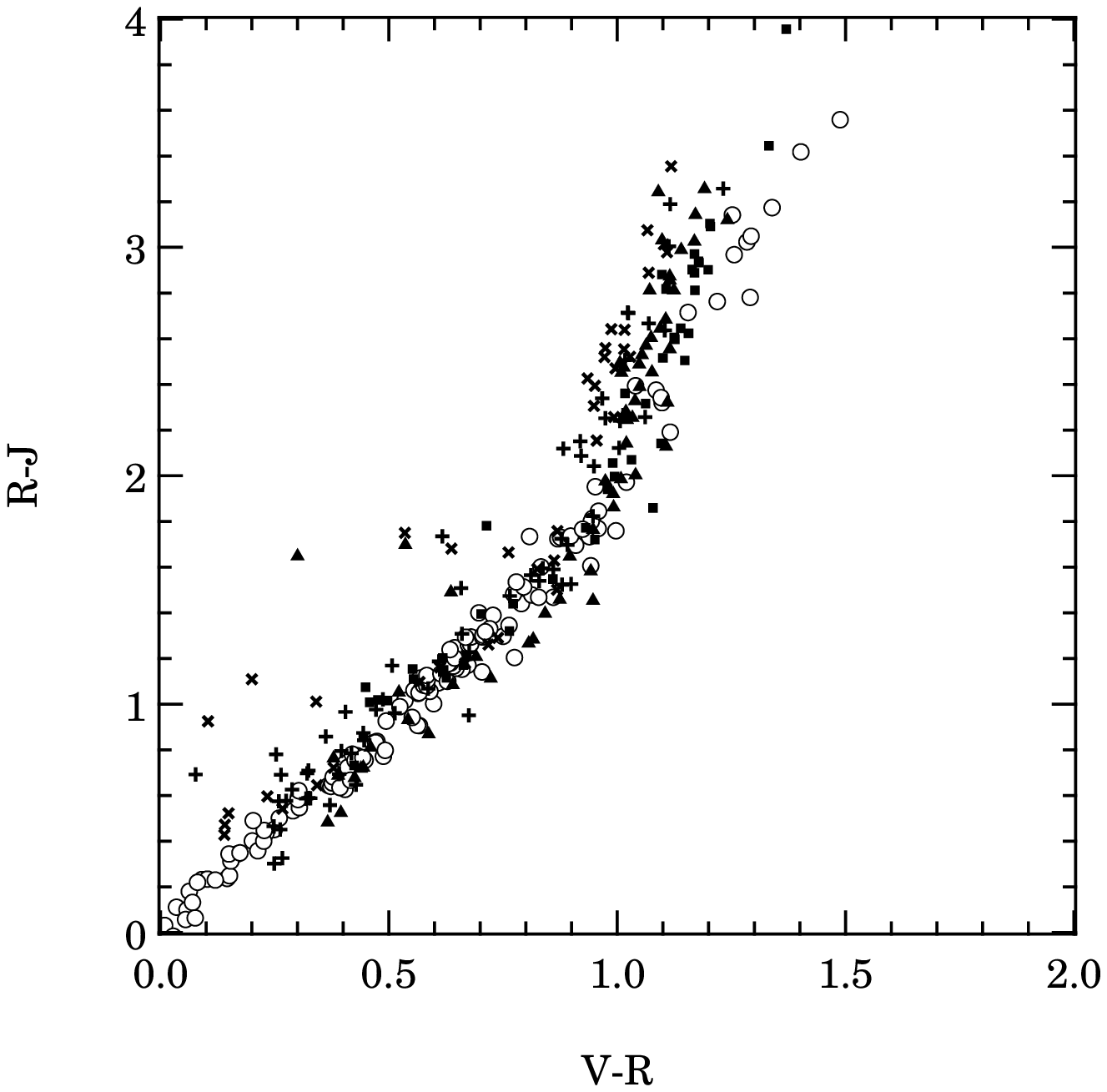,width=8.5cm}
    \epsfig{figure=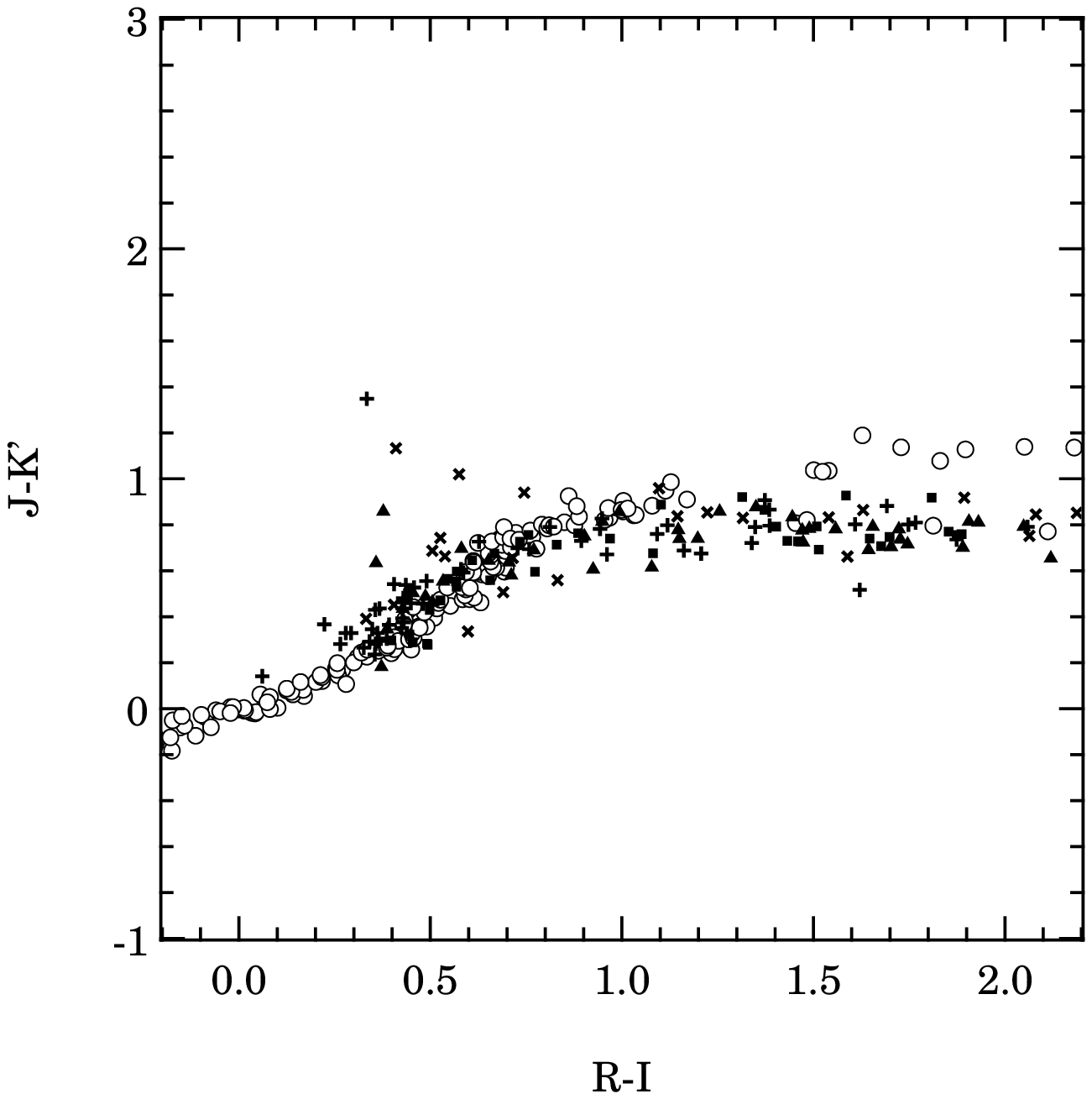,width=8.5cm}
  \caption{Comparison of $R\!-\!J$ vs. $V\!-\!R$ (upper panel) and 
    $J\!-\!K$ vs. $R\!-\!I$ (lower panel) colours derived by
    convolving stellar SEDs (see text) with the MUNICS filter curves
    (open circles) with a sample of bright stars ($R < 21$) identified
    in the MUNICS fields S2 f1--f4 (crosses), S4 f1--f4 (x), S6 f5--f8
    (squares), and S7 f5--f8 (triangles). }
  \label{f:mun_phot}
\end{figure}

The MUNICS imaging observations were carried out partly using
non-standard filters, or imperfect realisations of standard filters
(see Section~\ref{s:observations} below). Since the colours of objects
in the MUNICS catalogues extend to much redder colours than any
available photometric standard stars, we decided to work in the
MUNICS instrumental photometric system and not to transform magnitudes
into the standard Johnson-Kron-Cousins system. Linear transformation
to the Johnson-Kron-Cousins system would have caused magnitude errors
up to 1~mag, because the true transformations are highly non-linear,
especially for red objects. The MUNICS photometric zero-points are in
the Vega system.

Note that, since comparison of the object's colours with spectral
synthesis models is intended (e.g.\ for deriving photometric redshifts
or discussing the nature of EROs), it is important that the observed
colours and the synthetic colours are consistent with respect to the
filter set.

Accurate measurements of the transmission curves of the glass filters
and quantum efficiencies of the detectors were obtained and applied in
all subsequent synthetic photometry. The filter curves are shown in
Fig.~\ref{f:mun_filters}.

Such accurate knowledge of the filter system allows a reliable
calibration of different bands via colour--colour diagrammes of stars
which are compared with synthetic stellar sequences obtained from the
convolution of SEDs from stellar libraries with the transmission
curves. The absolute photometric zero-points can then be derived from
a single photometric observation in one band only.

Fig.~\ref{f:mun_phot} shows a comparison of $R\!-\!J$ vs. $V\!-\!R$
and $J\!-\!K$ vs. $R\!-\!I$ colours derived by convolving stellar SEDs
with the MUNICS filter curves with a sample of stars detected in the
MUNICS mosaic fields. The stellar SEDs used for computing the
synthetic colours are taken from the Bruzual-Persson-Gunn-Stryker
spectral library \cite{GS83,SEW79}, covering spectral types O5 to M8.
The agreement between the synthetic photometry and the data along the
stellar loci in colour--colour space demonstrates that the constructed
filter curves match the actual ones and that the photometric
zero-points are mutually consistent in the optical and the near-IR
regime (see also Sect.~\ref{s:optical} and Sect.~\ref{s:infrared}). It
is worth noting that we have no cool giants or supergiants in the
MUNICS sample (as those would occupy the redder sequence in $J\!-\!K$
at $R\!-\!I \ga 1.5$) and no stars of earlier type than roughly late F
to G.

%
%

\section{Observations and data reduction}
\label{s:observations}

\begin{table*}
 \begin{minipage}{140mm}
  \caption{The MUNICS mosaic and quasar fields}
  \label{t:fields}
  \begin{tabular}{lllllllll}
    Field & Mosaic/QSO & $z_{QSO}$ & $\alpha$ (J2000) & $\delta$ (J2000) & Filters & 
    $K'$ seeing & $E(B-V)$ & Remarks\\ 
    \hline
    S1 &f1--f2 & & 14:49:25 &   +65:55:31 & KJIRV & 1.01\arcsec & 0.017 & (1) \\
    S2 &f1--f4 & & 03:06:41 &   +00:01:12 & KJIRV & 1.30\arcsec & 0.080 & (3) \\
       &f5--f8 & & 03:06:41 & $-$00:13:30 & KJIRV & 1.17\arcsec & 0.083 & (3) \\
    S4 &f1--f4 & & 03:15:00 &   +00:07:41 & KJIRV & 0.95\arcsec & 0.094 & (3) \\
       &f5--f8 & & 03:14:05 &   +00:07:41 & KJIRV & 2.52\arcsec & 0.094 & (2) \\
    S3 &f1--f4 & & 09:04:38 &   +30:02:56 & KJIRV & 1.10\arcsec & 0.025 & (3) \\
       &f5--f8 & & 09:03:44 &   +30:02:56 & KJIRV & 1.11\arcsec & 0.027 & (3) \\
    S5 &f1--f4 & & 10:24:01 &   +39:46:37 & KJIRV & 1.17\arcsec & 0.012 & (3) \\
       &f5--f8 & & 10:25:14 &   +39:46:37 & KJIRV & 1.32\arcsec & 0.009 & (3) \\
    S6 &f1--f4 & & 11:55:58 &   +65:35:55 & KJIRV & 1.21\arcsec & 0.019 & (3) \\
       &f5--f8 & & 11:57:56 &   +65:35:55 & KJIRV & 1.41\arcsec & 0.015 & (3) \\
    S7 &f1--f4 & & 13:33:41 &   +16:51:44 & KJ    & 1.68\arcsec & 0.023 & (2) \\
       &f5--f8 & & 13:34:44 &   +16:51:44 & KJIRV & 1.12\arcsec & 0.029 & (3) \\
    \hline
    Q1  & J000701.3+002242 & 0.87 & 00:07:01 & +00:22:42 & KJIR  & 1.39\arcsec & 0.073 & PB 5741;       (4)\\ 
    Q2  & J000750.9+031733 & 1.10 & 00:07:51 & +03:17:32 & KJIR  & 1.07\arcsec & 0.020 & PB 5753;       (4)\\ 
    Q3  & J005444.0+144646 & 0.91 & 00:54:44 & +14:46:47 & KJIR  & 1.30\arcsec & 0.054 & PHL 892;       (4)\\ 
    Q4  & J005905.6+000651 & 0.72 & 00:59:06 & +00:06:52 & KJIR  & 1.00\arcsec & 0.027 & PHL 923           \\ 
    Q5  & J010026.8+043941 & 0.53 & 01:00:27 & +04:39:41 & KJIR  & 1.06\arcsec & 0.024 & UM 81;         (4)\\ 
    Q6  & J011033.7+015446 & 0.71 & 01:10:35 & +01:55:37 & KJIR  & 1.29\arcsec & 0.028 & MS 01080+0139; (4)\\ 
    Q7  & J011818.5+025806 & 0.67 & 01:18:19 & +02:58:06 & KJIR  & 1.23\arcsec & 0.039 & 3C 37             \\ 
    Q8  & J015838.9+034744 & 0.66 & 01:58:39 & +03:47:43 & KJIR  & 1.23\arcsec & 0.031 & UM 153;        (4)\\ 
    Q9  & J025937.5+003736 & 0.53 & 02:59:38 & +00:37:37 & KJIR  & 1.33\arcsec & 0.090 & US 3472           \\ 
    Q10 & J115517.9+653917 & 1.20 & 11:55:28 & +65:38:10 & KJIRV & 1.62\arcsec & 0.017 & 4C 65.13          \\
    Q11 & J122033.9+334312 & 1.51 & 12:20:34 & +33:43:10 & KJ    & 1.22\arcsec & 0.012 & 3C 270.1          \\ 
    Q12 & J133335.8+164904 & 2.08 & 13:33:40 & +16:48:14 & KJ    & 1.86\arcsec & 0.022 & PB 3977           \\
    Q13 & J133411.6+550125 & 1.25 & 13:34:12 & +55:01:25 & KJ    & 1.59\arcsec & 0.007 & 4C 55.27          \\ 
    Q14 & J135704.5+191906 & 0.71 & 13:57:05 & +19:19:07 & KJ    & 1.72\arcsec & 0.060 & PKS 1354+19       \\ 
    Q15 & J135817.6+575205 & 1.38 & 13:58:18 & +57:52:05 & KJ    & 1.89\arcsec & 0.010 & 4C 58.29          \\ 
    Q16 & J171938.4+480413 & 1.08 & 17:19:38 & +48:04:13 & KJ    & 1.79\arcsec & 0.019 & PG 1718+481       \\ 
    \hline
  \end{tabular}

  \medskip
  Field coordinates are given with respect to the image centres.\\
  QSO designations according to \citeN{VCV96}.\\

  (1) Mosaic incomplete in the near-IR. \\
  (2) Near-IR data quality poor.\\
  (3) Good data quality in all five wave-bands.\\
  (4) Radio quiet QSO.
\end{minipage}
\end{table*}


\subsection{Infrared observations and data reduction}
\label{s:infrared}

The $K'$-band and $J$-band imaging was obtained using the Omega-Prime
camera \cite{OmegaPrime98} at the prime focus of the Calar Alto 3.5-m
telescope. Omega-Prime is equipped with a HAWAII 1024$^2$ HgCdTe
array. The image scale is 0.396~arcsec per pixel, resulting in a $6.75
\arcmin \times 6.75\arcmin$ field of view. The $K'$ filter
($\lambda_0=2.12\mu$, $\Delta\lambda=0.35\mu$; see \citeNP{WC92}) was
used because it significantly reduces the thermal background seen by
the detector relative to the standard $K$ filter, thus gaining
sensitivity. Table \ref{t:runs} lists all observing runs undertaken to
present date.

The $K'$-band data were observed using a dithering pattern consisting
of 16 positions within an area of 30\arcsec$\times$ 30\arcsec laid out
on a $4 \times 4$ grid with 10\arcsec spacing between adjacent grid
points. The data were recorded using a randomized sequence of these 16
positions.

On each position 28 seconds of net exposure time were collected,
divided into several shorter exposures as necessary depending on the
ambient temperature and thus the level of the thermal background. The
length of the single exposures was always chosen such that
non-linearity of the detector was negligible.

This 16~position cycle was repeated 3 times yielding 48 frames and a
total exposure time of 1344~s. The $J$-band images were observed using
the same dithering pattern with longer integration times of 80~s on
each position and therefore needed only one cycle, giving a total of
1280~s.

The near-IR mosaic fields consist of four such Omega-Prime pointings
arranged in a $2 \times 2$ configuration with 6$\arcmin$ offset in
each direction measured from field centre to field centre. Each mosaic
then covers a total area of 162 square arc minutes, counting only the
central area with the longest total exposure time and removing
overlaps and borders due to the dithering pattern (see
Fig.~\ref{f:fgeom}).

On photometric nights, standard stars from the UKIRT Faint IR Standard
Stars catalogue \cite{UKIRTfaint92} were observed several times during
the night at different air masses to determine the photometric zero
point and the atmospheric extinction coefficient. To increase the
number of standard star measurements available for each night, the
calibrations of further stars in the UKIRT fields by
\citeN{NorthernJHK98} were included. Night-to-night variations in the
zero-point were typically less than 0.1~mag. Targets observed during
non-photometric nights were re-observed (with shorter exposure time)
at least once during photometric conditions to assure accurate
photometric calibration. The typical formal uncertainties in the
zero-points were $0.05$~mag in $K'$ and $0.06$~mag in $J$. The
extinction coefficients were found to be stable for all runs with
typical values around $0.08\pm0.025$~mag per airmass in $K'$ and
$0.12\pm0.02$~mag per airmass in $J$. By comparison with synthetic
photometry as explained in Sect.~\ref{s:photsys} we conclude that
additional systematic errors in the near-IR calibration as well as
systematic offsets between the near-IR and the optical wave-bands
cannot be larger than $\sim 0.1$~mag.

The data were reduced using standard image processing algorithms
within {\sc iraf}\footnote{{\sc iraf} is distributed by the National
  Optical Astronomy Observatories, which are operated by the
  Association of Universities for Research in Astronomy, Inc., under
  cooperative agreement with the National Science Foundation.}. For
each frame a sky frame was constructed from typically 6 to 12
(temporally) adjacent frames where bright objects and detector defects
have been masked out, and which were scaled to have the same median
counts. These frames were then median-combined using clipping to
suppress fainter sources and otherwise deviant pixels to produce a sky
frame. The sky frame was scaled to the median counts of each image
before subtraction to account for variations of sky brightness on
short time-scales. The sky-subtracted images were flat-fielded using
dome flats to remove pixel-to-pixel fluctuations in quantum
efficiency. The frames were then registered to high accuracy using the
brightest $\sim 10$ objects and finally co-added, again using clipping
to suppress highly deviant pixels due to cosmic ray events and
defective pixels on the array, after being scaled to airmass zero and
to a common photometric zero-point.

The $2\times 2$ mosaic images were produced by registering the images
using objects in the overlap regions, simultaneously cross checking
the photometric calibration. Before combining, the images were
adjusted to have the same background counts computed from the mode of
the pixel values in `empty' sky regions of the images to correct for
residual differences in sky brightness. Absolute astrometric
calibration of the images is discussed in Sect.~\ref{s:astrometry}
below.

\begin{table}
  \caption{MUNICS observing runs}
  \label{t:runs}
  \begin{tabular}{llll}
    Date & Tel. & Instrument & Remarks \\ 
    \hline
    1996 24-27.10 & CA35  & $\Omega'$ & Quasar fields \\
    1997 15-19.5  & CA35  & $\Omega'$ & Quasar fields \\
    1998 8-14.4   & CA35  & $\Omega'$ & \\
    1998 12-17.5  & CA35  & $\Omega'$ & \\
    1998 28.5-1.6 & CA22  & CAFOS & \\
    1998 16-18.11 & McD27 & IGI & Quasar fields\\
    1998 16-20.12 & CA22  & CAFOS & \\
    1998 23-30.12 & CA35  & $\Omega'$ & \\
    1999 18.3     & Wdst  & MONICA & Calibrations \\
    1999 27.5-3.6 & CA35  & $\Omega'$ & \\
    1999 9-18.6   & CA22  & CAFOS & \\
    2000 26-31.5  & CA35  & MOSCA & Spectroscopy \\
    2000 27-28.5  & HET   & LRS   & Spectroscopy \\
    2000 16.7     & CA35  & $\Omega'$ & (1) \\
    2000 20-22.11 & ESO-VLT & FORS1/2 & Spectroscopy \\
    2000 24-28.11 & CA35  & MOSCA & Spectroscopy \\
    2000 5.12     & CA22  & CAFOS & (1) \\
    2000 17-18.12 & CA22  & CAFOS & (1) \\
    2000 19.12    & CA35  & $\Omega'$ & (1) \\
    2001 17-21.1  & CA35  & MOSCA & Spectroscopy \\
    2001 11-13.2  & CA35  & $\Omega'$ & (1) \\ \hline
  \end{tabular}

  \medskip
  (1) Re-imaging of fields with poor data quality. \\

  CA22 and CA35 are the 2.2-m telescope and the 3.5-m telescope of Calar Alto 
  Observatory, respectively. McD27 is the 2.7-m telescope and HET is 
  the Hobby-Eberly Telescope, both of McDonald Observatory, Austin, Texas. 
  Wdst is the 0.8-m telescope of Wendelstein 
  Observatory operated by the Universit\"ats-Sternwarte M\"unchen. \\
\end{table}


\subsection{Optical observations and data reduction}
\label{s:optical}

Optical imaging of the mosaic fields was performed at the Calar Alto
2.2-m telescope in the $V$, $R$, and $I$ bands using the Calar Alto
Faint Object Spectrograph (CAFOS) focal reducer in direct imaging
mode. CAFOS was equipped with a SITe 2048$^2$ CCD detector, yielding a
resolution of 0.53 arcsec per pixel and a circular field of view (due
to vignetting by optics) of $16\arcmin$ in diameter. The $V$-band filter
used was a standard Johnson filter, the $R$-band filter was an $R_2$
filter ($\lambda_0=0.648\mu, \Delta\lambda=0.168\mu$), slightly
narrower and bluer than Kron-Cousins $R$. The $I$-band filter was an
RG780 filter with the red cutoff set by the CCD (see
Fig.~\ref{f:mun_filters}.) Total exposure times were 2700~s in $V$ and
$I$, and 1800~s in $R$, divided into several shorter exposures taken
with offsets of $\sim 15\arcsec$, depending on the presence of bright
stars and on seeing conditions to avoid too many saturated objects.

The quasar fields were imaged using the Imaging Grism Instrument (IGI)
at the 2.7-m telescope of McDonald Observatory, using a 1024$^2$ TK4
CCD ($7\arcmin$ field of view) and Mould $R$ and $I$ interference
filters. Exposure times 1800~s in $R$ and 2700~s in $I$, again divided
into several shorter exposures.

The optical CCD data were reduced in a fairly standard manner using
{\sc iraf}, except for cosmic ray cleaning. The frames were
bias/overscan corrected and then flat-fielded using a combination of
dome flats and sky flats. The $I$-band frames showed considerable
fringing. Fringe images were created from the affected series of
science exposures and occasionally also from twilight flats by
medianing de-registered images after masking bright sources by hand as
necessary. In some cases it was necessary to subtract a low order fit
to the overall background in the science frames prior to construction
of the fringe image to account for changes in the illumination pattern
present in the images in the case where a bright star was close to the
image border. The fringe images were then appropriately scaled and
subtracted from the affected frames.

Cosmic ray events were identified by searching for narrow local maxima
in the image and fitting a bivariate rotated Gaussian to each maximum.
A locally deviant pixel is then replaced by the mean value of the
surrounding pixels if the Gaussian obeys appropriate flux ratio and
sharpness criteria \cite{GRF00}. Such a procedure is much more
expensive in terms of computing time (roughly 10 CPU minutes per
frame) compared to standard median filtering techniques, but is much
more reliable in finding cosmic ray events in the wings of objects and
in cleaning long cosmic ray trails.

The re-imaging system of CAFOS causes substantial radial distortion of
the image which had to be dealt with before co-adding the offset
images.  Therefore the frames were rectified using the known
distortion equation, a polynomial of fourth order in the distance from
the optical axis (K.\ Meisenheimer, private communication).

If necessary, variations in the background intensity across the frames
caused by scattered light were fitted and subtracted in each
individual frame. The images were then corrected for atmospheric
extinction and scaled to a common photometric zero-point before
finally being added using the positions of $\sim 15$ bright objects
for determination of the offsets between the individual frames.

During photometric nights, photometric standard stars were observed
(\citeNP{Lando92}; \citeNP{CABBHMS85}) and programme fields with
insecure calibrations were re-observed with short exposures. The run
at the Wendelstein 0.8-m telescope was devoted to such re-calibration
to have independent calibrations for the fields. For each field, a
photometric zero point and the atmospheric extinction were determined.
No colour terms were fitted to the calibration data, as explained in
Sect.~\ref{s:photsys}. The typical formal uncertainties in the
photometric calibration were $\sim 0.08$~mag in $I$, $\sim 0.04$~mag
in $R$, and $\sim 0.05$~mag in $V$. The extinction coefficients were
usually consistent with a Rayleigh atmosphere, with a few nights
showing higher extinction, albeit within the variations typical for
Calar Alto.  As with the near-IR observations, consistency was checked
against synthetic colours of stars, again finding no systematic
offsets. We conclude that systematic errors in the optical photometric
calibration are again smaller than $0.1$~mag in all filters.


\subsection{Astrometry}
\label{s:astrometry}

Astrometric solutions were computed for all $K'$-band images to
translate pixel coordinates into celestial coordinates. For this
purpose, astrometric standards from the USNO-SA1.0 catalogue
\cite{USNO} were selected in each frame. The celestial coordinates of
these stars were matched against the pixel position using the {\sc
  iraf} task {\sc ccxymatch}, and the plate solution was computed
using {\sc ccmap}. The typical scatter is less than 0.4 arcsec rms.

The $V$, $R$, $I$, and $J$ images of each field were registered against
the $K'$-band image by matching the positions of \mbox{$\sim 200$}
bright homogeneously distributed objects in the frames and determining
the coordinate transform from the $K'$-band system to each image in
the other four pass-bands using the tasks {\sc xyxymatch} and {\sc
  geomap} within {\sc iraf}. The scatter in the determined solutions
is less than 0.1 pixels rms in the transformation from $K'$ to $J$,
and less than 0.2 pixels rms from $K'$ to the optical frames. Note
that the frames themselves are not transformed. We only determine
accurate transformations and apply these later to the apertures in the
photometry process.

\subsection{Spectroscopy}
\label{s:spectroscopy}

A spectroscopic follow-up programme is currently being conducted at
the Calar-Alto 3.5-m telescope, the Hobby-Eberly Telescope, and the
VLT, aiming ultimately at a magnitude-limited redshift survey of the
$K'$-band selected catalogue. The results of these observations will
be discussed in a future paper. In the mean time, the spectroscopic
redshifts are used to calibrate photometric redshifts.

%
%

\section{Data analysis}
\label{s:analysis}

The construction of photometric catalogues from the reduced images
will be discussed in this chapter. The individual steps in this
process will be described in some detail. These include the detection
of objects in the $K'$-band images (see Sect.~\ref{s:detection}), a
first analysis of the survey's completeness
(Sect.~\ref{s:completeness}), the photometry of objects in all filters
(see Sect.~\ref{s:photometry}), and the separation of stars and
galaxies in the catalogue (see Sect.~\ref{s:classification}). The
chapter will be concluded by a comparison of number counts for
galaxies with previous studies (Sect.~\ref{s:numbercounts}) as a
further consistency check on our data set.


\subsection{Object detection}
\label{s:detection}

\begin{figure*}
  \centerline{\epsfig{figure=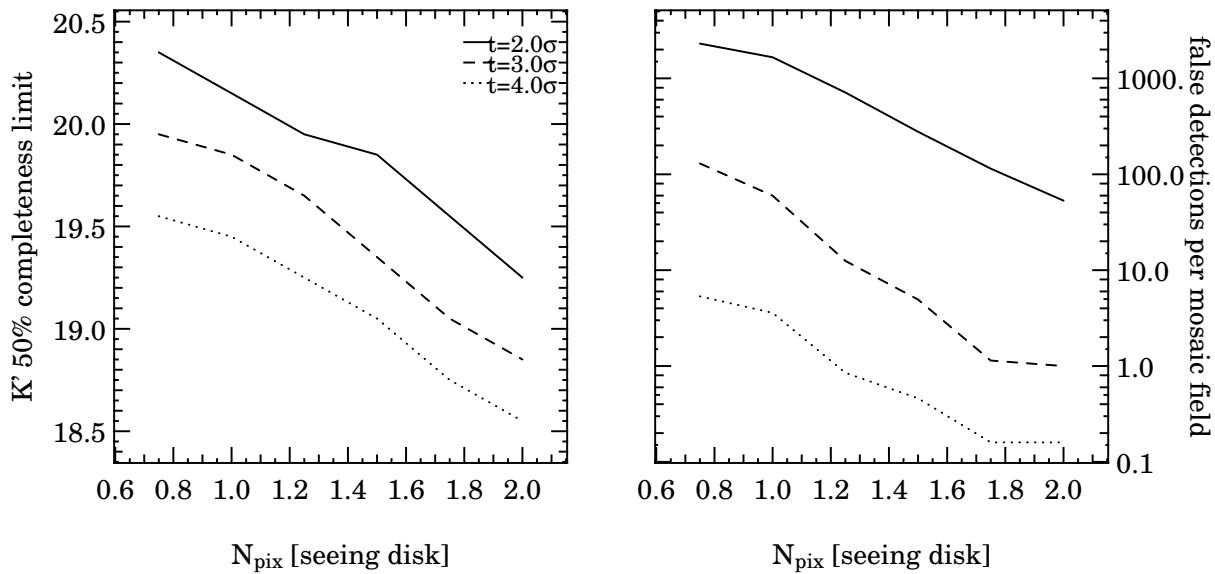,angle=-90,width=16cm}}
  \caption{The behaviour of the 50 per cent completeness 
    limit for point-like sources and the number of spurious source
    detections as a function of the detection threshold $t$ in units
    of the local background rms $\sigma$ and the required number
    $N_{pix}$ of consecutive pixels above the threshold in units of
    the seeing disk area $\pi ({\mathrm FWHM}/2)^2$. The {\it left
      panel} shows the change in limiting magnitude at 50 per cent
    completeness as a function of $N_{pix}$ for detection thresholds
    of 2.0$\sigma$ (solid line), 3.0$\sigma$ (dashed line), and
    4.0$\sigma$ (dotted line). The {\it right panel} shows the number
    of spurious sources integrated over all magnitudes per image (one
    mosaic field), again as a function of $N_{pix}$ and the detection
    threshold $t$. Line styles as in the left panel.}
  \label{f:false_detect}
\end{figure*}

Object detection was performed using the {\sc yoda} source extraction
software \cite{YODA00}. This package was specifically designed to be
used in multi-band imaging surveys, where the background noise is
often inhomogeneous across the images -- in mosaiced frames or in
dithered images where the exposure time is a function of position --
and where the frames do not share a common coordinate system and pixel
scale, due to the use of multiple telescopes and imagers. The second
point was considered a serious problem since re-sampling the images to
a common coordinate system introduces quite considerable noise for
faint sources.

Sources are detected by requiring a minimum number $N_{pix}$ of
consecutive pixels to lie above a certain threshold $t$ expressed in
units of the local rms $\sigma$ of the background noise. To foster
detection of faint sources, the images are convolved with a Gaussian
of FWHM equal to the seeing in the image. The choice of the number of
consecutive pixels $N_{pix}$ and the threshold $t$ is somewhat a
tradeoff between limiting magnitude at some completeness fraction, say
50 per cent, and the number of tolerable spurious detections per unit
image area \cite{Saha95}. Note that, since we aim at purely $K'$-band
selected catalogues for most of our applications, the presence or
absence of a source in other wave-bands cannot be used for
confirmation or rejection of sources, so the expected number of
spurious detections per unit image area is of great interest to us.

To find reasonable values for $N_{pix}$ and $t$ we performed
simulations on the $K'$-band image of one of our mosaic fields (S6
f5--f8). The dependence of the 50 per cent completeness limit on
$N_{pix}$ and $t$ was determined by adding point sources to the
$K'$-band image and recording the fraction of the objects recovered by
the detection software as a function of $N_{pix}$ and $t$. The number
of false detections was determined by looking for positive detections
in an inverted (multiplied by $-1$) version of the image, after
convincing ourselves that the background noise was sufficiently well
approximated by a Gaussian. Fig.~\ref{f:false_detect} shows the
results of these tests.

At the depth of our data we detect roughly 1000 objects per mosaic
field. Accepting 1 per cent contamination by false detections, i.e.\ 
roughly 10 false objects per mosaic field we fixed the detection
threshold at $t=3\sigma$ and the minimum number of consecutive pixels
at 1.4 times the seeing disk area, $N_{pix}=1.4 \pi ({\mathrm
  FWHM}/2)^2$, (10 pixels at 1\arcsec, 16 pixels at 1.5\arcsec seeing
for the near-IR frames) and performed object detection using these
parameters on all $K'$-band images.


\subsection{Completeness}
\label{s:completeness}

\begin{table*}
  \begin{minipage}{130mm}
    \caption{Completeness limits for the MUNICS mosaic fields}
    \label{t:comptable}
    \begin{tabular}{lllllllllllll}
      & \multicolumn{2}{c}{S2 f1--f4} & \multicolumn{2}{c}{S4 f1--f4} & 
      \multicolumn{2}{c}{S5 f1--f4} & \multicolumn{2}{c}{S6 f1--f4} &
      \multicolumn{2}{c}{S6 f5--f8} & \multicolumn{2}{c}{S7 f5--f8} \\
      Band & 90\% & 50\%  &  90\% & 50\%  &  90\% & 50\%   &  90\% & 50\% 
      &  90\% & 50\%   &  90\% & 50\% \\
      \hline
      K &  19.05 & 19.45 & 18.69 & 19.18 & 19.42 & 19.80 & 
           18.26 & 18.56 & 19.53 & 19.92 & 19.23 & 19.83\\
      J &  20.29 & 20.68 & 20.62 & 21.04 & 20.53 & 20.93 & 
           20.06 & 20.44 & 20.69 & 21.06 & 20.84 & 21.39\\
      I &  21.86 & 22.29 & 22.18 & 22.60 & 21.72 & 22.07 & 
           22.35 & 22.72 & 21.70 & 22.08 & 21.94 & 22.34\\
      R &  22.78 & 23.21 & 23.18 & 23.50 & 22.94 & 23.32 & 
           23.25 & 23.58 & 22.95 & 23.37 & 22.88 & 23.28\\
      V &  23.13 & 23.51 & 23.62 & 23.94 & 23.36 & 23.72 & 
           23.36 & 23.68 & 23.50 & 23.89 & 22.96 & 23.28\\
      \hline
    \end{tabular}
  \end{minipage}
\end{table*}

\begin{figure}
  \centerline{\epsfig{figure=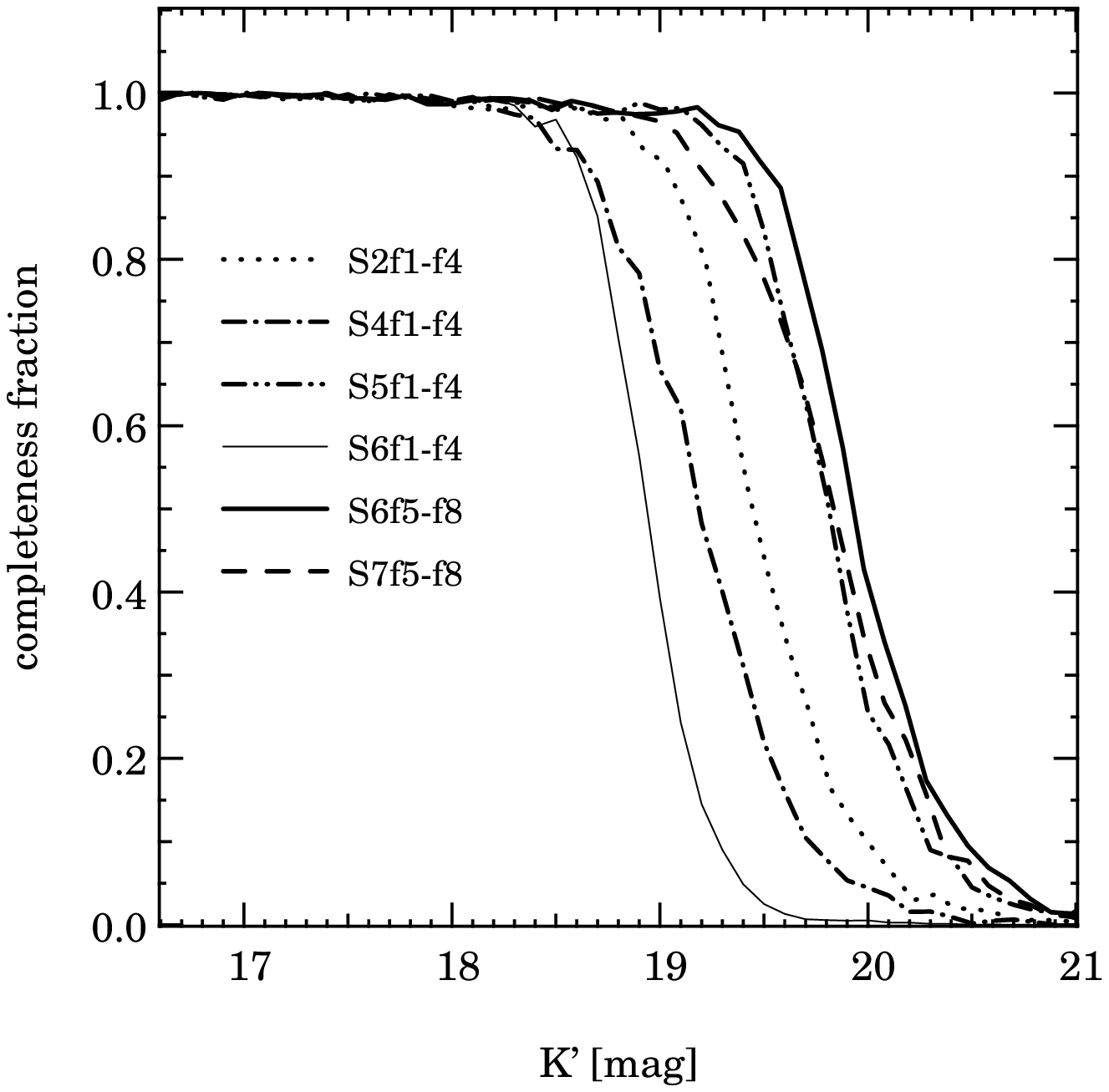,width=8.5cm}}
  \caption{Completeness fraction of point-like sources 
    as a function of magnitude in the $K'$-band mosaic fields as
    determined by adding Moffat-type objects to the images and
    recording the fraction of such objects that was recovered by the
    detection process.}
  \label{f:compl_k}
\end{figure}

To estimate the completeness limits of the MUNICS $K'$-selected
catalogues, Monte--Carlo simulations were carried out to determine the
detection completeness as a function of magnitude in each $K'$-band
image.

In these simulations, $250$ artificial objects with a Moffat-type PSF
having the same FWHM as stars in the frames were added to the MUNICS
images. For this purpose the {\sc iraf} package {\sc artdata} was
used. A constant distribution of apparent magnitudes of the artificial
objects was applied. Then the object detection algorithm was run on
these frames, and the number of re-detected objects was recorded as a
function of magnitude. This procedure was repeated $250$ times in
order to decrease statistical errors.

Fig.~\ref{f:compl_k} shows the results of these simulations for 6 of
the mosaic fields where imaging data in all five colours are
available.  The 50 per cent and 90 per cent completeness limits of the
$K'$-band images are listed in Table~\ref{t:comptable}. From these
simulations we conclude that the mosaic fields comprise a reasonably
homogeneous data set, with field-to-field variations in 50 per cent
completeness of order $\sim 0.4$~mag, with the exception of S6 f1--f4
which is considerably shallower. The quasar fields, which are not
shown here, have completeness levels in the same range as the shown
mosaic fields.

Simulations with extended objects having de Vaucouleurs and
exponential surface brightness profiles have also been performed. In
general they yield 50 per cent completeness limits between 0.5 and 1
magnitude brighter than those determined for stellar sources,
depending mostly on the adopted profile scale-length. The full results
of these simulations will be extensively discussed in a future paper
in the context of surface-brightness selection effects.


\subsection{Photometry}
\label{s:photometry}

\begin{figure}
  \centerline{\epsfig{figure=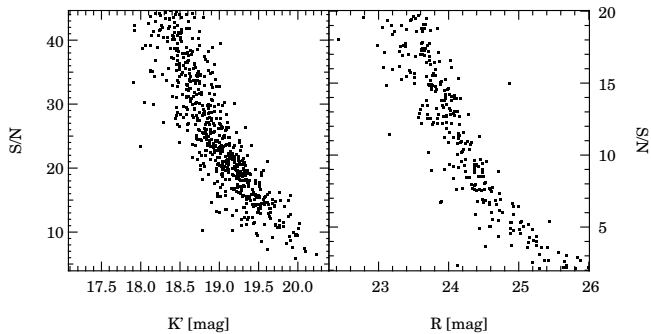,angle=-90, width=8.5cm}}
  \caption{Signal-to-noise ratio in $K'$ (left panel) and $R$ (right panel) 
    as a function of magnitude in circular apertures of 5\arcsec
    diameter for objects taken from the $K'$-selected catalogue of the
    field {\mbox S6 f5--f8}. The signal-to-noise ratio is defined here
    as the signal-to-noise ratio of the aperture photometry, i.e.\ 
    total (sky-subtracted) flux within the aperture divided by the
    total noise within the aperture, with contributions to the latter
    coming from Poisson fluctuations in the object as well as the
    background, and the error in the determination of the background.}
  \label{f:sn}
\end{figure}

\begin{figure}
  \centerline{\epsfig{figure=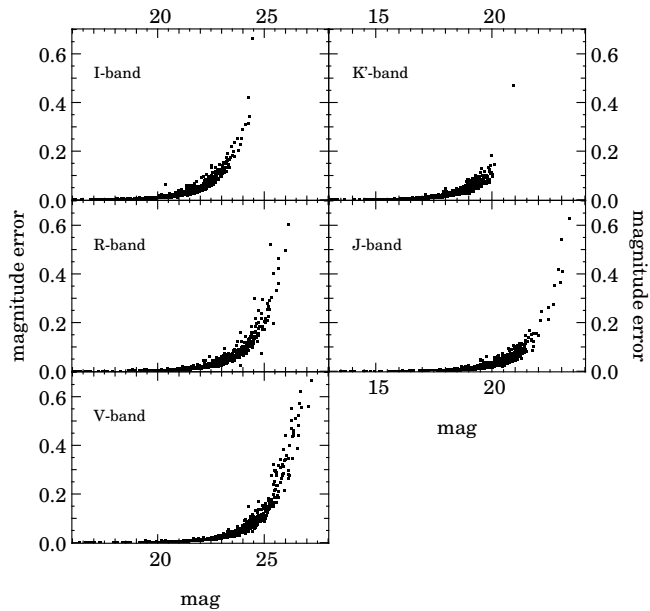,width=8.5cm}}
  \caption{Magnitude errors as a function of magnitude in the
    mosaic field S6 f5--f8 in $K', J, I, R$, and $V$ as measured in
    circular apertures of 5\arcsec diameter for $K'$-band selected
    objects.  Aperture fluxes are measured in every pass-band for each
    object present in the $K'$-band catalogue irrespective of a
    detection in any other band.}
  \label{f:magerr}
\end{figure}

Photometry was performed in elliptical apertures the shape of which
were determined from the first and second moments of the light
distribution in the $K'$-band image, as described in \citeN{YODA00},
and additionally in fixed size circular apertures of 5 and 7 arc
seconds diameter. To ensure measurement at equal physical scales in
every pass-band, the individual frames were convolved to the same
seeing FWHM, namely that of the image with the worst seeing in each
field. The signal-to-noise ratio as a function of magnitude for the
5\arcsec apertures is shown in Fig.~\ref{f:sn}. At the 50 per cent
completeness limit in the $K'$ band (19.59), the signal-to-noise ratio
is $\sim 10$. For such an object having an $R\!-\!K'$ colour of 6, the
signal-to-noise ratio in the $R$ band is roughly 3.

Aperture fluxes and magnitudes were computed for each object present
in the $K'$-band catalogue irrespective of a detection in any other
band. For this purpose the centroid coordinates of the sources
detected in the $K'$-band images were transformed to the other frames
using the full astrometric transformations as determined in
Sect.~\ref{s:astrometry}. The shape of the apertures were transformed
using only the linear terms of the transformation.

The photometric accuracy for the 5\arcsec aperture magnitudes is
roughly 0.1~mag at $K'=19$~mag. This error estimate includes the
effects of photon noise and uncertainty in background determination
and subtraction, but does not include (systematic) errors due to the
photometric calibration. Fig.~\ref{f:magerr} shows plots of the
magnitude error vs. object magnitude for one mosaic field in all five
pass-bands.


\subsection{Star-galaxy separation}
\label{s:classification}

Star--galaxy separation relies on {\sc yoda}'s image classification
stage which is based on a Bayesian analysis of the probability that an
object's light distribution is due to an unresolved (point-like)
source by comparison with light distributions constructed from the
image's PSF. {\sc yoda}'s classification parameters are calculated for
all objects in the catalogue in all available pass-bands.

As demonstrated in \citeN{YODA00}, classification is reliable across
wave-bands and imaging instruments, and stellar sources almost do not
scatter out of the stellar locus in parameter space (except in the
presence of crowding).  Rather, images of faint galaxies as they
become smaller at larger distances, move onto the locus of point-like
sources.

Using the multi-pass-band information available in MUNICS allows us to
push the limit of reliable classification by using for each object the
classification information in those pass-bands where the
signal-to-noise ratio is highest. Therefore, in the mosaic fields
where 5 colours are available, we classify as stellar every source
that is classified as stellar by {\sc yoda} in the three pass-bands
with highest signal-to-noise. In the quasar fields, where less colour
information is available, we rely on the two images with highest
signal-to-noise.

As can be seen in Fig.~\ref{f:data_colours}, objects classified as
stars occupy the clearly defined stellar sequence in the $R\!-\!J$ vs.
$J\!-\!K'$ colour-colour plane, with only very few objects classified
as stellar having a $J\!-\!K'$ colour redder than $\sim 1$. These are
either misclassified faint and compact galaxies or very late-type
stars or brown dwarfs, the latter is a possibility for those objects
having also red $R\!-\!J$ colour. The objects lying on the stellar
sequence at $R\!-\!J \ga 2$ and which are classified as galaxies were
found to be faint and barely resolved objects failing the
classification as a star only due to their appearance in one filter.
In many cases an obvious reason -- like a second close object -- could
be identified. We conclude that most of these objects are, in fact,
misclassified stars. The total fraction of point-like sources in the
catalogues is $\sim 10$ per cent.

We have also checked the results of the image-based classification
against spectral classification for those objects where spectroscopy
was already available, namely 45 galaxies and 53 stars having
$R<20.5$.  All these objects were correctly classified.


\subsection{Galactic extinction}
\label{s:galextinction}

We use the Galactic reddening maps provided by \citeN{SFD98} using a
value of $R_V=3.1$ to calculate $A_\lambda=R_\lambda E(B-V)$ to
correct the measured magnitudes for Galactic foreground extinction.
The values of $E(B-V)$ for our fields are given in
Table~\ref{t:fields}.


\subsection{Galaxy number counts}
\label{s:numbercounts}

\begin{figure}
  \centerline{\epsfig{figure=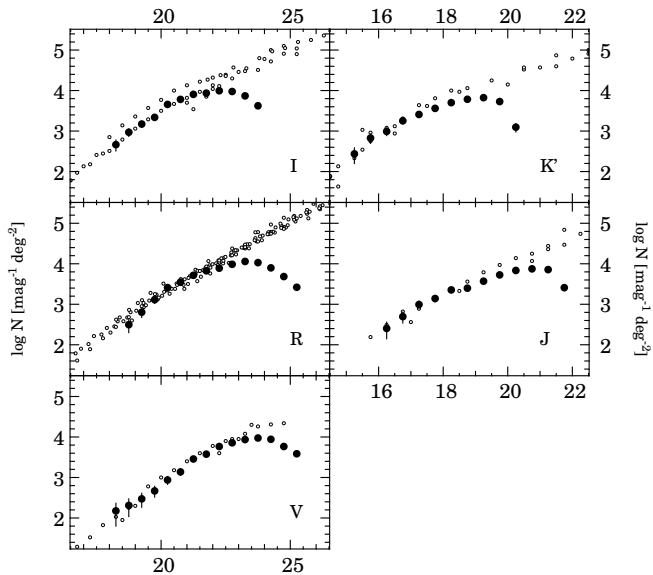,width=8.5cm}}
  \caption{Number counts for galaxies in $K'$, $J$, $I$, $R$, and $V$
    from MUNICS data (filled circles) and previous studies (open
    circles), as described in the text. The counts shown are average
    number counts from all available MUNICS data and have not been
    corrected for incompletness. Error bars indicate Poisson errors.}
  \label{f:nc}
\end{figure}

In Table~\ref{t:nc} we present number counts of galaxies in the MUNICS
mosaic fields in all five filters $K'$, $J$, $I$, $R$, and $V$. These
counts are also shown in fig.~\ref{f:nc}, together with a compilation
of number counts from the literature. Object catalogues were generated
independently for each pass-band for this purpose, and star--galaxy
separation is based on the PSF classification as described above,
using only single pass-band information. The data were not transformed
into the standard magnitude system for this comaprison. Completeness
corrections were not applied to these galaxy number counts, but
corrected counts will be presented in the context of a more detailed
completeness analysis in a future paper.  The counts are average
counts from all the available MUNICS mosaic fields, with
field-to-field variations in the number counts being on the level of
0.1~dex. The errors given in Table~\ref{t:nc} only include Poissonian
errors.

The number counts are compared to the following literature values.
\citeNP{GCW93,CGHSHW94,GPCM94,Detal95,GSCF96} for the $K$ band,
\citeNP{Saracco99,Teplitz99} for the $J$ band,
\citeNP{Tyson88,LCG91,CRGINOW95,GSCF96,HDF96,HCL98} for the $I$ band,
\citeNP{CN84,HM84,IPQ86,Koo86,SSF86,YG87,Tyson88,JFSEP91,MSFJ91,P91,CJB93,SH93,DPDMD94,MFS95,MSFR95,SHYC95,MSCFG96,BD97,HPMCBSS97,ADCZFG99,MSCMF00}
for the $R$ band, and \citeNP{DPDMD94,CRGINOW95,GSCF96} for the $V$
band. 

We generally find good agreement with previously published number
counts in all pass-bands, again as a consistency check confirming
the quality of our photometry.

\begin{table*}
  \caption{$K', J, I, R$, and $V$-band galaxy number counts for the MUNICS 
    mosaic fields. The counts as a function magnitude and the error of 
    the counts are given in logarithmic units. The values have not been 
    corrected for incompleteness. The errors are Poissonian errors only.}
  \label{t:nc}
  \begin{tabular}{llllllllllllllll}
 & \multicolumn{3}{c}{$K'$} & \multicolumn{3}{c}{$J$}
 & \multicolumn{3}{c}{$I$} & \multicolumn{3}{c}{$R$}  
 & \multicolumn{3}{c}{$V$} \\  
 $mag$ & $\log n$ & $\sigma_{low}$ & $\sigma_{high}$ &
 $\log n$ & $\sigma_{low}$ & $\sigma_{high}$ &
 $\log n$ & $\sigma_{low}$ & $\sigma_{high}$ &
 $\log n$ & $\sigma_{low}$ & $\sigma_{high}$ &
 $\log n$ & $\sigma_{low}$ & $\sigma_{high}$ \\
 \hline
14.25 & 1.19 & 0.47 & 1.37 \\
14.75 & 2.11 & 1.38 & 2.32 \\
15.25 & 2.44 & 2.20 & 2.59 \\
15.75 & 2.83 & 2.70 & 2.93 \\
16.25 & 2.99 & 2.89 & 3.07 & 2.40 & 2.15 & 2.56\\
16.75 & 3.25 & 3.18 & 3.32 & 2.69 & 2.54 & 2.81& 2.14 & 1.61 & 2.34 \\                                         
17.25 & 3.41 & 3.35 & 3.46 & 2.99 & 2.89 & 3.08& 2.13 & 1.40 & 2.33 \\                                         
17.75 & 3.56 & 3.51 & 3.61 & 3.14 & 3.06 & 3.21& 2.38 & 2.15 & 2.53 \\                                         
18.25 & 3.70 & 3.66 & 3.74 & 3.35 & 3.29 & 3.41& 2.66 & 2.50 & 2.77 & 1.92 & 1.72 & 2.04 & 2.17 & 1.80 & 2.36\\
18.75 & 3.79 & 3.75 & 3.82 & 3.40 & 3.34 & 3.45& 2.97 & 2.86 & 3.05 & 2.49 & 2.29 & 2.63 & 2.30 & 2.03 & 2.47\\
19.25 & 3.83 & 3.79 & 3.86 & 3.57 & 3.52 & 3.61& 3.17 & 3.09 & 3.24 & 2.80 & 2.67 & 2.90 & 2.47 & 2.26 & 2.61\\
19.75 & 3.73 & 3.69 & 3.77 & 3.73 & 3.68 & 3.76& 3.33 & 3.27 & 3.39 & 3.11 & 3.02 & 3.18 & 2.66 & 2.51 & 2.78\\
20.25 & 3.10 & 2.98 & 3.18 & 3.84 & 3.80 & 3.87& 3.65 & 3.61 & 3.69 & 3.41 & 3.35 & 3.46 & 2.93 & 2.83 & 3.02\\
20.75 &      &      &      & 3.87 & 3.84 & 3.91& 3.78 & 3.74 & 3.81 & 3.54 & 3.49 & 3.58 & 3.13 & 3.05 & 3.20\\
21.25 &      &      &      & 3.86 & 3.82 & 3.89& 3.90 & 3.87 & 3.93 & 3.71 & 3.67 & 3.75 & 3.45 & 3.40 & 3.50\\
21.75 &      &      &      & 3.41 & 3.34 & 3.47& 3.93 & 3.90 & 3.96 & 3.82 & 3.79 & 3.85 & 3.57 & 3.52 & 3.61\\
22.25 &      &      &      &      &      &     & 3.99 & 3.96 & 4.02 & 3.89 & 3.85 & 3.92 & 3.76 & 3.72 & 3.80\\
22.75 &      &      &      &      &      &     & 3.97 & 3.94 & 4.00 & 3.98 & 3.95 & 4.01 & 3.85 & 3.82 & 3.88\\
23.25 &      &      &      &      &      &     & 3.86 & 3.83 & 3.90 & 4.05 & 4.03 & 4.08 & 3.93 & 3.90 & 3.96\\
23.75 &      &      &      &      &      &     & 3.62 & 3.57 & 3.66 & 4.02 & 4.00 & 4.05 & 3.97 & 3.94 & 4.00\\
24.25 &      &      &      &      &      &     & 3.36 & 3.29 & 3.41 & 3.89 & 3.86 & 3.92 & 3.94 & 3.91 & 3.97\\
24.75 &      &      &      &      &      &     & 2.75 & 2.58 & 2.86 & 3.68 & 3.64 & 3.72 & 3.76 & 3.72 & 3.80\\
25.25 &      &      &      &      &      &     &      &      &      & 3.42 & 3.36 & 3.47 & 3.58 & 3.53 & 3.62\\
25.75 &      &      &      &      &      &     &      &      &      & 3.03 & 2.93 & 3.11 & 3.16 & 3.06 & 3.23\\
\hline
  \end{tabular}
\end{table*}

\subsection{Colour distributions and objects at $z \ge 1$}

\begin{figure*}
  \begin{minipage}{\textwidth}
    \epsfig{figure=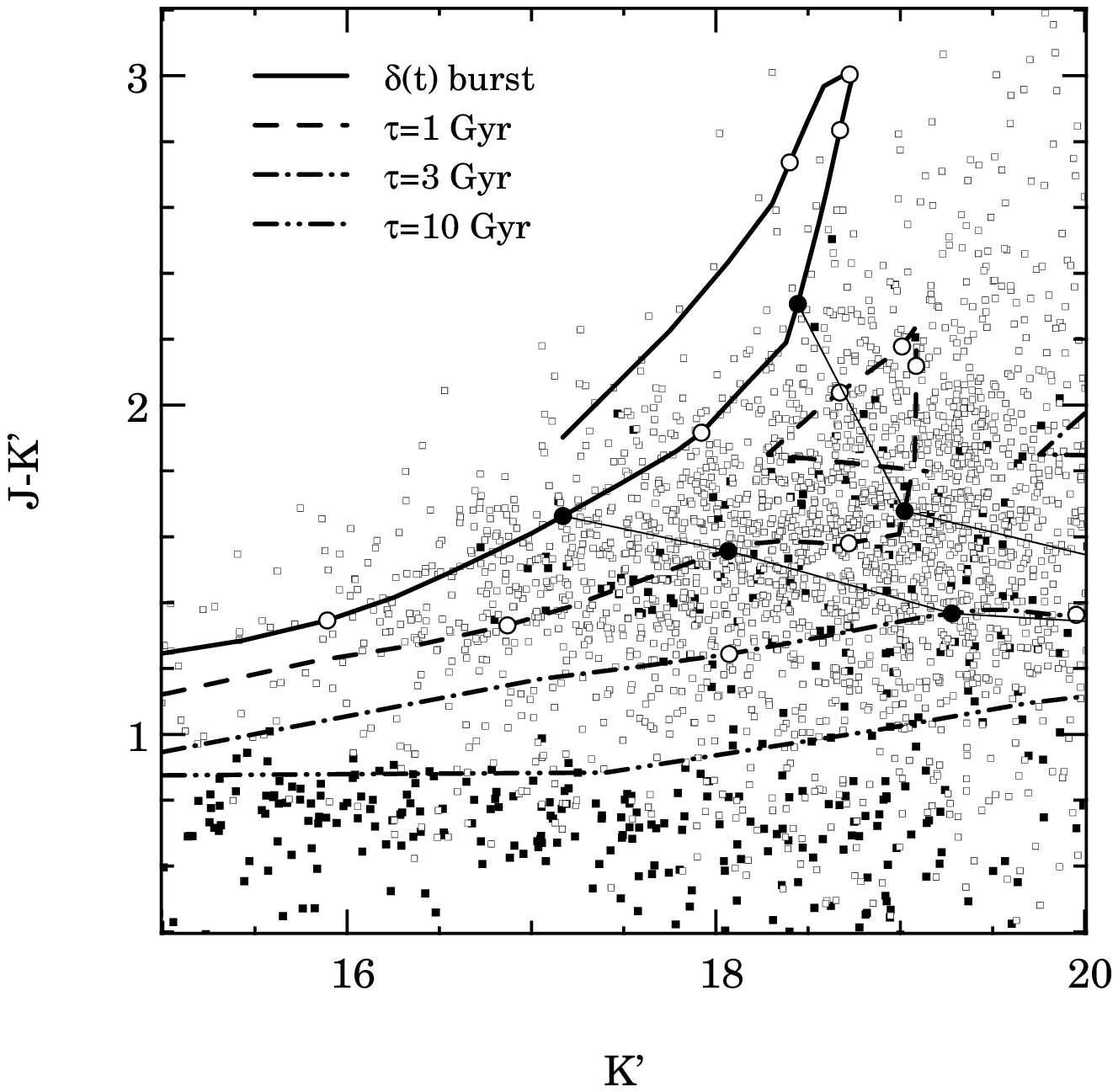,width=0.48\textwidth}
    \epsfig{figure=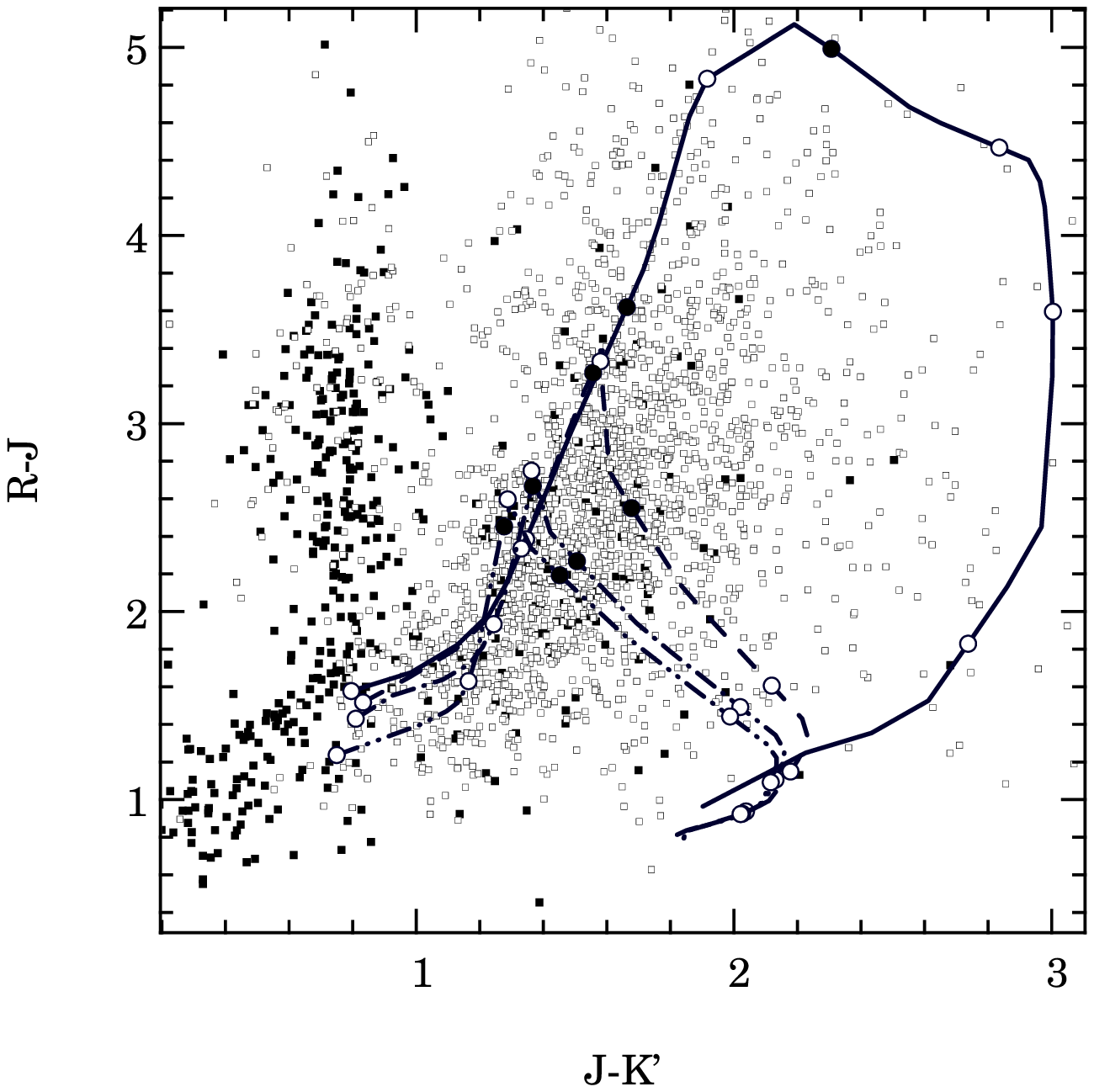,width=0.48\textwidth}\vspace*{1cm}
    \epsfig{figure=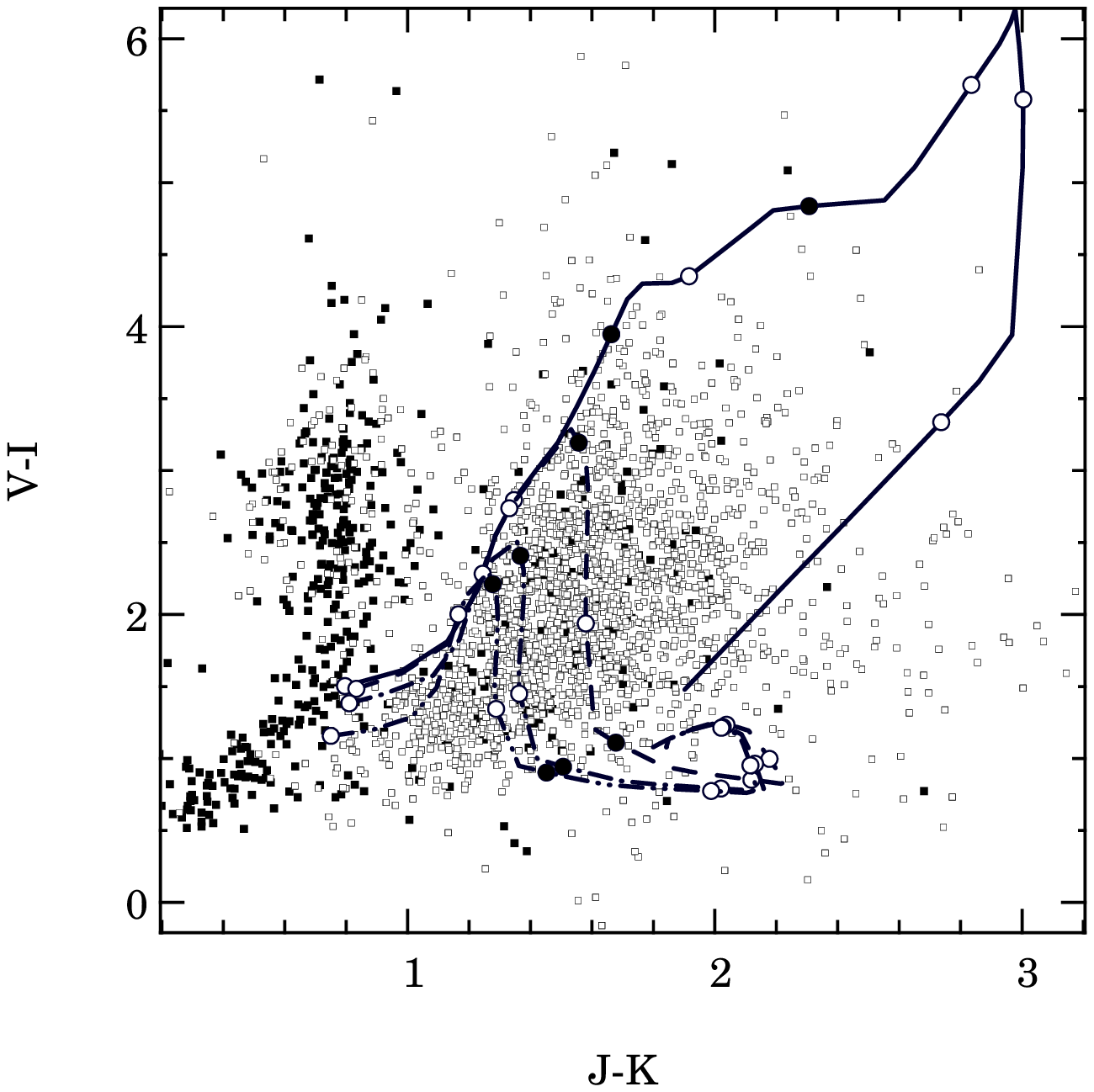,width=0.48\textwidth}\hfill
    \epsfig{figure=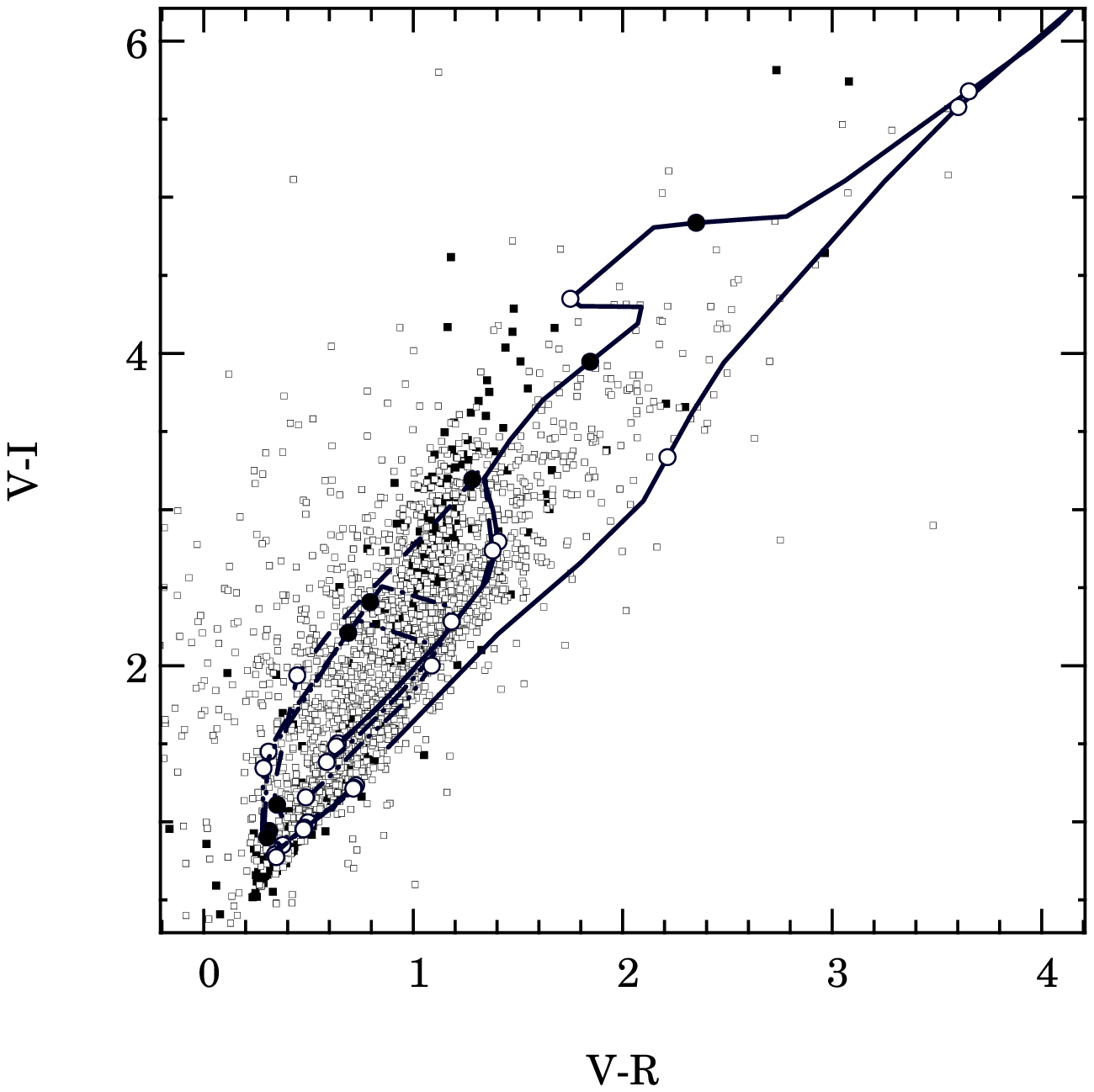,width=0.48\textwidth}
  \caption{%
    Colour--magnitude and colour--colour diagrammes
    for MUNICS objects taken from 3 mosaic fields (S2 f1--f4, S6
    f5--f8, and S7 f5--f8) containing 2977 sources. Objects classified
    as stellar are marked with filled squares, extended objects are
    marked with open squares.  Also shown are stellar population
    synthesis models for different star-formation histories.  The
    model parameters are the same as in Fig.~\ref{f:models_kjk}.
    Redshift along the model tracks is marked by circles at a $z$
    spacing of 0.5, with $z=1$ and $z=2$ being accentuated by filled
    circles. The lines of constant redshift at $z=1$ and $z=2$ are
    drawn as thin solid lines.}
  \label{f:data_colours}
  \end{minipage}
\end{figure*}

In Fig.\ref{f:data_colours} we show the $J\!-\!K'$ vs. $K'$
colour--magnitude diagramme and the $R\!-\!J$ vs. $J\!-\!K'$,
$V\!-\!I$ vs. $J\!-\!K'$, and $V\!-\!I$ vs. $V\!-\!R$ colour--colour
diagrammes for MUNICS data from three mosaic fields. The total number
of objects shown is 2977, of which 286 are classified as point-like.
These plots also contain the tracks defined by the stellar population
synthesis models described in detail in Sect.~\ref{s:sensitivity}.
Briefly, the models are an SSP, and three exponential star formation
histories with $e$-folding times of 1, 3, and 10~Gyr forming at $z=4$.
The models have been normalised such that they represent typical $L^*$
objects at $z=0$, with $L^*$ chosen according to their `photometric'
Hubble type. The cosmology adopted is again $H_0=65, \Omega_0=0.3,
\Omega_\Lambda=0.7$.

These models reasonably envelope the region in the colour--magnitude
$J\!-\!K'$ vs. $K'$ plane occupied by the data, with the SSP model
following the outline of the data points along the bright and red edge
as might be expected since any further star formation or a later
formation epoch would render the object bluer relative to the SSP.

It is also worth noting that the models constitute a continuous
sequence with the duration of the star formation as the parameter in
the $R\!-\!J$ vs. $J\!-\!K'$ plane, closely following the SSP track up
to a redshift of $\sim 1$, then rapidly turning bluer in $R\!-\!J$
while still getting redder in $J\!-\!K'$. A significant fraction of
objects between the SSP and the 1~Gyr track is compatible with being
well evolved objects at a redshift $z \ga 1$. How many objects exactly
populate this region is an important question which will be addressed
in a future paper.

We finally conclude from these diagrammes that the quality of our data
meets the requirements expressed in Sect.~\ref{s:concept} and that we
are in a position to construct a catalogue containing a large number
of massive field galaxies in the redshift range $0.5 \la z \la 1.5$ to
study their evolution in detail.

\section{Summary}

The Munich Near-IR Cluster Survey (MUNICS) is a wide-area,
medium-deep, photometric survey selected in the $K'$ band. It covers
an area of roughly one square degree in the $K'$ and $J$ near-IR
pass-bands with additional complementary optical imaging in the $V$,
$R$, and $I$ bands.

MUNICS has been undertaken to study the evolution of both field
galaxies and galaxy clusters out to redshifts around unity, and to
investigate the nature of extremely red objects and their connection
to the population of massive field spheroidal galaxies.

The survey area consists of $16$ $6\arcmin \times 6\arcmin$ fields
targeted at QSOs with redshifts $0.5 < z < 2$ and $7$ $28\arcmin
\times 13\arcmin$ stripes targeted at `random' high Galactic latitude
fields. Ten of the QSO fields were additionally imaged in $R$ and $I$,
and $0.6~\mathrm{deg}^2$ of the randomly selected fields were imaged
in the $V$, $R$, and $I$ bands. The resulting object catalogues were
strictly selected in $K'$, having a limiting magnitude ($50$ per cent
completeness) of $K' \sim 19.5$~mag and $J \sim 21$~mag,
sufficiently deep to detect passively evolving early-type systems up
to a redshift of $z \la 1.5$ and luminosity of $0.5 L^*$.  The optical
data reach a depth of roughly $R \sim 23.5$~mag. The project's main
scientific aims are the identification of galaxy clusters at redshifts
around unity and the selection of a large sample of field early-type
galaxies at $0 < z < 1.5$ for evolutionary studies.

In this paper we describe the selection of survey fields as well as
the observations and the reduction of the near-infrared and optical
data.  We define our photometric system and show it to be internally
consistent by checking it against synthetic photometry of stars from
stellar libraries. The construction of the $K'$-selected object
catalogue is described in detail, particularly the choice of
parameters for object detection, which ensures completeness to as
faint magnitudes as possible while keeping the rate of false
detections in our strictly $K'$-selected catalogue small. Photometry
of the objects in the catalogue is performed in elliptical apertures
on frames convolved to the same PSF in all filters, in order to
guarantee measurements of the flux in equal physical areas. Stars and
galaxies in the fields are classified using a Bayesian analysis of the
light distribution in the images.

The quality of the survey data in terms of signal-to-noise ratio and
limiting magnitude is discussed, with the completeness of the survey
fields being characterised by Monte--Carlo simulations of point
sources in the MUNICS frames. Also, galaxy number counts are presented
in the five filters $K'$, $J$, $I$, $R$, and $V$ and compared to
counts published by other authors.

Finally, we show the $J\!-\!K'$ vs. $K'$ colour--magnitude diagramme
and the $R\!-\!J$ vs. $J\!-\!K'$, $V\!-\!I$ vs. $J\!-\!K'$, and
$V\!-\!I$ vs. $V\!-\!R$ colour--colour diagrammes for MUNICS objects,
together with stellar population-synthesis models for different
star-formation histories and conclude that the data set presented is
suitable for extracting a catalogue of massive field galaxies in the
redshift range $0.5 \la z \la 1.5$ for evolutionary studies and
follow-up observations.

\section*{Acknowledgements}

The authors would like to thank the staff at Calar Alto Observatory
and McDonald Observatory for their extensive support during the many
observing runs of this project. G.\ J.\ Hill is especially
acknowledged for his commitment at the McDonald Observatory
facilities. The authors also thank Claus G\"ossl for supplying his
cosmics filter, Nigel Metcalfe for making number count data available
in electronic form, and Klaus Meisenheimer for providing the
distortion correction for CAFOS. This research has made use of NASA's
Astrophysics Data System (ADS) Abstract Service and the NASA/IPAC
Extragalactic Database (NED).  The MUNICS project was supported by the
Deutsche Forschungsgemeinschaft, {\it Sonderforschungsbereich 375,
  Astroteilchenphysik}. CMdO is grateful to the Alexander von Humboldt
Foundation for financial support provided during a visit to the
Universit\"ats-Sternwarte M\"unchen from Sep. 1997 to Aug. 1998.

\bibliography{mnrasmnemonic,literature} \bibliographystyle{mnras}

\begin{thebibliography}{}

\bibitem[\protect\citeauthoryear{{Arnouts} et~al.}{{Arnouts}
  et~al.}{1999}]{ADCZFG99}
{Arnouts} S., {D'Odorico} S., {Cristiani} S., {Zaggia} S., {Fontana} A.,
  {Giallongo} E., 1999, A\&A, 341, 641

\bibitem[\protect\citeauthoryear{{Bahcall} \& {Fan}}{{Bahcall} \&
  {Fan}}{1998}]{BF98}
{Bahcall} N.~A.,  {Fan} X., 1998, ApJ, 504, 1

\bibitem[\protect\citeauthoryear{{Bahcall}, {Fan}, \& {Cen}}{{Bahcall}
  et~al.}{1997}]{BFC97}
{Bahcall} N.~A., {Fan} X.,  {Cen} R., 1997, ApJ, 485, L53

\bibitem[\protect\citeauthoryear{{Baum}}{{Baum}}{1962}]{Baum62}
{Baum} W.~A., 1962, in IAU Symp. 15: Problems of Extra-Galactic Research,
  Vol.~15, p. 390

\bibitem[\protect\citeauthoryear{{Ben{\'\i}tez}}{{Ben{\'\i}tez}}{2000}]{Benite%
z00}
{Ben{\'\i}tez} N., 2000, ApJ, 536, 571

\bibitem[\protect\citeauthoryear{{Bertin} \& {Dennefeld}}{{Bertin} \&
  {Dennefeld}}{1997}]{BD97}
{Bertin} E.,  {Dennefeld} M., 1997, A\&A, 317, 43

\bibitem[\protect\citeauthoryear{{Bizenberger} et~al.}{{Bizenberger}
  et~al.}{1998}]{OmegaPrime98}
{Bizenberger} P., {McCaughrean} M.~J., {Birk} C., {Thompson} D.,  {Storz} C.,
  1998, Proceedings of SPIE, 3354, 825

\bibitem[\protect\citeauthoryear{{Brinchmann} \& {Ellis}}{{Brinchmann} \&
  {Ellis}}{2000}]{BE00}
{Brinchmann} J.,  {Ellis} R.~S., 2000, ApJ, 536, L77

\bibitem[\protect\citeauthoryear{{Broadhurst}, {Ellis}, \&
  {Shanks}}{{Broadhurst} et~al.}{1988}]{BES88}
{Broadhurst} T.~J., {Ellis} R.~S.,  {Shanks} T., 1988, MNRAS, 235, 827

\bibitem[\protect\citeauthoryear{Bruzual \& Charlot}{Bruzual \&
  Charlot}{1993}]{BC93}
Bruzual G.~A.,  Charlot S., 1993, ApJ, 405, 538

\bibitem[\protect\citeauthoryear{{Casali} \& {Hawarden}}{{Casali} \&
  {Hawarden}}{1992}]{UKIRTfaint92}
{Casali} M.~M.,  {Hawarden} T., 1992, {JCMT} {UKIRT} Newsletter, 4, 33

\bibitem[\protect\citeauthoryear{{Casertano} et~al.}{{Casertano}
  et~al.}{1995}]{CRGINOW95}
{Casertano} S., {Ratnatunga} K.~U., {Griffiths} R.~E., {Im} M., {Neuschaefer}
  L.~W., {Ostrander} E.~J.,  {Windhorst} R.~A., 1995, ApJ, 453, 599

\bibitem[\protect\citeauthoryear{Christian et~al.}{Christian
  et~al.}{1985}]{CABBHMS85}
Christian C.~A., Adams M., Barnes J.~V., Butcher H., Hayes D.~S., Mould J.~R.,
  Siegel M., 1985, PASP, 97, 363

\bibitem[\protect\citeauthoryear{{Cimatti} et~al.}{{Cimatti}
  et~al.}{1999}]{CDSPetal99}
{Cimatti} A. et~al., 1999, A\&A, 352, L45

\bibitem[\protect\citeauthoryear{{Colless} et~al.}{{Colless}
  et~al.}{1990}]{LDSS90}
{Colless} M., {Ellis} R.~S., {Taylor} K.,  {Hook} R.~N., 1990, MNRAS, 244, 408

\bibitem[\protect\citeauthoryear{{Couch}, {Jurcevic}, \& {Boyle}}{{Couch}
  et~al.}{1993}]{CJB93}
{Couch} W.~J., {Jurcevic} J.~S.,  {Boyle} B.~J., 1993, MNRAS, 260, 241

\bibitem[\protect\citeauthoryear{{Couch} \& {Newell}}{{Couch} \&
  {Newell}}{1984}]{CN84}
{Couch} W.~J.,  {Newell} E.~B., 1984, ApJS, 56, 143

\bibitem[\protect\citeauthoryear{{Cowie} et~al.}{{Cowie}
  et~al.}{1994}]{CGHSHW94}
{Cowie} L.~L., {Gardner} J.~P., {Hu} E.~M., {Songaila} A., {Hodapp} K.-W.,
  {Wainscoat} R.~J., 1994, ApJ, 434, 114

\bibitem[\protect\citeauthoryear{{de Propris} et~al.}{{de Propris}
  et~al.}{1999}]{PSEDE99}
{de Propris} R., {Stanford} S.~A., {Eisenhardt} P.~R., {Dickinson} M.,
  {Elston} R., 1999, AJ, 118, 719

\bibitem[\protect\citeauthoryear{{Djorgovski} et~al.}{{Djorgovski}
  et~al.}{1995}]{Detal95}
{Djorgovski} S. et~al., 1995, ApJ, 438, 1

\bibitem[\protect\citeauthoryear{{Driver} et~al.}{{Driver}
  et~al.}{1994}]{DPDMD94}
{Driver} S.~P., {Phillipps} S., {Davies} J.~I., {Morgan} I.,  {Disney} M.~J.,
  1994, MNRAS, 266, 155

\bibitem[\protect\citeauthoryear{{Drory}}{{Drory}}{2001}]{YODA00}
{Drory} N., 2001, A\&A, submitted

\bibitem[\protect\citeauthoryear{{Eke}, {Cole}, \& {Frenk}}{{Eke}
  et~al.}{1996}]{ECF96}
{Eke} V.~R., {Cole} S.,  {Frenk} C.~S., 1996, MNRAS, 282, 263

\bibitem[\protect\citeauthoryear{{Eke} et~al.}{{Eke} et~al.}{1998}]{ECFH98}
{Eke} V.~R., {Cole} S., {Frenk} C.~S.,  {Patrick Henry} J., 1998, MNRAS, 298,
  1145

\bibitem[\protect\citeauthoryear{{Elston}, {Rieke}, \& {Rieke}}{{Elston}
  et~al.}{1988}]{ERR88}
{Elston} R., {Rieke} G.~H.,  {Rieke} M.~J., 1988, ApJ, 331, L77

\bibitem[\protect\citeauthoryear{{Fern{\'a}ndez-Soto}, {Lanzetta}, \&
  {Yahil}}{{Fern{\'a}ndez-Soto} et~al.}{1999}]{FLY99}
{Fern{\'a}ndez-Soto} A., {Lanzetta} K.~M.,  {Yahil} A., 1999, ApJ, 513, 34

\bibitem[\protect\citeauthoryear{{Gardner}, {Cowie}, \& {Wainscoat}}{{Gardner}
  et~al.}{1993}]{GCW93}
{Gardner} J.~P., {Cowie} L.~L.,  {Wainscoat} R.~J., 1993, ApJ, 415, L9

\bibitem[\protect\citeauthoryear{{Gardner} et~al.}{{Gardner}
  et~al.}{1996}]{GSCF96}
{Gardner} J.~P., {Sharples} R.~M., {Carrasco} B.~E.,  {Frenk} C.~S., 1996,
  MNRAS, 282, L1

\bibitem[\protect\citeauthoryear{{Gardner} et~al.}{{Gardner}
  et~al.}{1997}]{GSFC97}
{Gardner} J.~P., {Sharples} R.~M., {Frenk} C.~S.,  {Carrasco} B.~E., 1997, ApJ,
  480, L99

\bibitem[\protect\citeauthoryear{{Glazebrook} et~al.}{{Glazebrook}
  et~al.}{1994}]{GPCM94}
{Glazebrook} K., {Peacock} J.~A., {Collins} C.~A.,  {Miller} L., 1994, MNRAS,
  266, 65

\bibitem[\protect\citeauthoryear{{G\"ossl}, {Riffeser}, \& {Fliri}}{{G\"ossl}
  et~al.}{2001}]{GRF00}
{G\"ossl} C.~A., {Riffeser} A.,  {Fliri} J., 2001, A\&A, in preparation

\bibitem[\protect\citeauthoryear{{Gunn} \& {Stryker}}{{Gunn} \&
  {Stryker}}{1983}]{GS83}
{Gunn} J.~E.,  {Stryker} L.~L., 1983, ApJS, 52, 121

\bibitem[\protect\citeauthoryear{{Hall} \& {Mackay}}{{Hall} \&
  {Mackay}}{1984}]{HM84}
{Hall} P.,  {Mackay} C.~D., 1984, MNRAS, 210, 979

\bibitem[\protect\citeauthoryear{{Hogg} et~al.}{{Hogg}
  et~al.}{1997}]{HPMCBSS97}
{Hogg} D.~W., {Pahre} M.~A., {McCarthy} J.~K., {Cohen} J.~G., {Blandford} R.,
  {Smail} I.,  {Soifer} B.~T., 1997, MNRAS, 288, 404

\bibitem[\protect\citeauthoryear{{Hu} \& {Ridgway}}{{Hu} \&
  {Ridgway}}{1994}]{HR94}
{Hu} E.~M.,  {Ridgway} S.~E., 1994, AJ, 107, 1303

\bibitem[\protect\citeauthoryear{{Huang}, {Cowie}, \& {Luppino}}{{Huang}
  et~al.}{1998}]{HCL98}
{Huang} J., {Cowie} L.~L.,  {Luppino} G.~A., 1998, ApJ, 496, 31

\bibitem[\protect\citeauthoryear{{Hunt} et~al.}{{Hunt}
  et~al.}{1998}]{NorthernJHK98}
{Hunt} L.~K., {Mannucci} F., {Testi} L., {Migliorini} S., {Stanga} R.~M.,
  {Baffa} C., {Lisi} F.,  {Vanzi} L., 1998, AJ, 115, 2594

\bibitem[\protect\citeauthoryear{{Infante}, {Pritchet}, \&
  {Quintana}}{{Infante} et~al.}{1986}]{IPQ86}
{Infante} L., {Pritchet} C.,  {Quintana} H., 1986, AJ, 91, 217

\bibitem[\protect\citeauthoryear{{Jones} et~al.}{{Jones}
  et~al.}{1991}]{JFSEP91}
{Jones} L.~R., {Fong} R., {Shanks} T., {Ellis} R.~S.,  {Peterson} B.~A., 1991,
  MNRAS, 249, 481

\bibitem[\protect\citeauthoryear{{Kauffmann} \& {Charlot}}{{Kauffmann} \&
  {Charlot}}{1998}]{KC98a}
{Kauffmann} G.,  {Charlot} S., 1998, MNRAS, 297, L23

\bibitem[\protect\citeauthoryear{{Koo}}{{Koo}}{1985}]{Koo85}
{Koo} D.~C., 1985, AJ, 90, 418

\bibitem[\protect\citeauthoryear{{Koo}}{{Koo}}{1986}]{Koo86}
{Koo} D.~C., 1986, ApJ, 311, 651

\bibitem[\protect\citeauthoryear{{Landolt}}{{Landolt}}{1992}]{Lando92}
{Landolt} A.~U., 1992, AJ, 104, 340

\bibitem[\protect\citeauthoryear{{Lilly}, {Cowie}, \& {Gardner}}{{Lilly}
  et~al.}{1991}]{LCG91}
{Lilly} S.~J., {Cowie} L.~L.,  {Gardner} J.~P., 1991, ApJ, 369, 79

\bibitem[\protect\citeauthoryear{{Lilly} et~al.}{{Lilly}
  et~al.}{1995a}]{CFRS95}
{Lilly} S.~J., {Le Fevre} O., {Crampton} D., {Hammer} F.,  {Tresse} L., 1995a,
  ApJ, 455, 50

\bibitem[\protect\citeauthoryear{{Lilly} et~al.}{{Lilly}
  et~al.}{1995b}]{CFRS695}
{Lilly} S.~J., {Tresse} L., {Hammer} F., {Crampton} D.,  {Le Fevre} O., 1995b,
  ApJ, 455, 108

\bibitem[\protect\citeauthoryear{{Maraston}}{{Maraston}}{1998}]{Maraston98}
{Maraston} C., 1998, MNRAS, 300, 872

\bibitem[\protect\citeauthoryear{{McLeod} et~al.}{{McLeod}
  et~al.}{1995}]{MBRTF95}
{McLeod} B.~A., {Bernstein} G.~M., {Rieke} M.~J., {Tollestrup} E.~V.,  {Fazio}
  G.~G., 1995, ApJS, 96, 117

\bibitem[\protect\citeauthoryear{{Metcalfe}, {Fong}, \& {Shanks}}{{Metcalfe}
  et~al.}{1995}]{MFS95}
{Metcalfe} N., {Fong} R.,  {Shanks} T., 1995, MNRAS, 274, 769

\bibitem[\protect\citeauthoryear{{Metcalfe} et~al.}{{Metcalfe}
  et~al.}{1996}]{MSCFG96}
{Metcalfe} N., {Shanks} T., {Campos} A., {Fong} R.,  {Gardner} J.~P., 1996,
  Nat, 383, 236

\bibitem[\protect\citeauthoryear{{Metcalfe} et~al.}{{Metcalfe}
  et~al.}{2001}]{MSCMF00}
{Metcalfe} N., {Shanks} T., {Campos} A., {McCracken} H.~J.,  {Fong} R., 2001,
  MNRAS, in press

\bibitem[\protect\citeauthoryear{{Metcalfe} et~al.}{{Metcalfe}
  et~al.}{1991}]{MSFJ91}
{Metcalfe} N., {Shanks} T., {Fong} R.,  {Jones} L.~R., 1991, MNRAS, 249, 498

\bibitem[\protect\citeauthoryear{{Metcalfe} et~al.}{{Metcalfe}
  et~al.}{1995}]{MSFR95}
{Metcalfe} N., {Shanks} T., {Fong} R.,  {Roche} N., 1995, MNRAS, 273, 257

\bibitem[\protect\citeauthoryear{{Mobasher}, {Sharples}, \& {Ellis}}{{Mobasher}
  et~al.}{1993}]{MSE93}
{Mobasher} B., {Sharples} R.~M.,  {Ellis} R.~S., 1993, MNRAS, 263, 560

\bibitem[\protect\citeauthoryear{{Monet} et~al.}{{Monet} et~al.}{1996}]{USNO}
{Monet} D. et~al., 1996, "USNO-SA1.0".
\newblock U.S. Naval Observatory, Washington DC, 1996.

\bibitem[\protect\citeauthoryear{{Picard}}{{Picard}}{1991}]{P91}
{Picard} A., 1991, AJ, 102, 445

\bibitem[\protect\citeauthoryear{{Rix} \& {Rieke}}{{Rix} \&
  {Rieke}}{1993}]{RR93}
{Rix} H.,  {Rieke} M.~J., 1993, ApJ, 418, 123

\bibitem[\protect\citeauthoryear{{Saha}}{{Saha}}{1995}]{Saha95}
{Saha} P., 1995, AJ, 110, 916

\bibitem[\protect\citeauthoryear{{Sandage}, {Binggeli}, \& {Tammann}}{{Sandage}
  et~al.}{1985}]{SBT85}
{Sandage} A., {Binggeli} B.,  {Tammann} G.~A., 1985, AJ, 90, 1759

\bibitem[\protect\citeauthoryear{{Saracco} et~al.}{{Saracco}
  et~al.}{1999}]{Saracco99}
{Saracco} P., {D'Odorico} S., {Moorwood} A., {Buzzoni} A., {Cuby} J.~.,
  {Lidman} C., 1999, A\&A, 349, 751

\bibitem[\protect\citeauthoryear{{Saracco} et~al.}{{Saracco}
  et~al.}{1997}]{SIGM97}
{Saracco} P., {Iovino} A., {Garilli} B., {Maccagni} D.,  {Chincarini} G., 1997,
  AJ, 114, 887

\bibitem[\protect\citeauthoryear{{Schlegel}, {Finkbeiner}, \&
  {Davis}}{{Schlegel} et~al.}{1998}]{SFD98}
{Schlegel} D.~J., {Finkbeiner} D.~P.,  {Davis} M., 1998, ApJ, 500, 525

\bibitem[\protect\citeauthoryear{{Smail} et~al.}{{Smail} et~al.}{1995}]{SHYC95}
{Smail} I., {Hogg} D.~W., {Yan} L.,  {Cohen} J.~G., 1995, ApJ, 449, L105

\bibitem[\protect\citeauthoryear{{Smail} et~al.}{{Smail}
  et~al.}{1999}]{SIKetal99}
{Smail} I., {Ivison} R.~J., {Kneib} J.-P., {Cowie} L.~L., {Blain} A.~W.,
  {Barger} A.~J., {Owen} F.~N.,  {Morrison} G., 1999, MNRAS, 308, 1061

\bibitem[\protect\citeauthoryear{{Stanford}, {Eisenhardt}, \&
  {Dickinson}}{{Stanford} et~al.}{1998}]{SED98}
{Stanford} S.~A., {Eisenhardt} P.~R.,  {Dickinson} M., 1998, ApJ, 492, 461

\bibitem[\protect\citeauthoryear{{Stanford}, {Eisenhardt}, \&
  {Dickinson}}{{Stanford} et~al.}{1995}]{SED95}
{Stanford} S.~A., {Eisenhardt} P.~R.~M.,  {Dickinson} M., 1995, ApJ, 450, 512

\bibitem[\protect\citeauthoryear{{Stanford} et~al.}{{Stanford}
  et~al.}{1997}]{SEESSD97}
{Stanford} S.~A., {Elston} R., {Eisenhardt} P.~R., {Spinrad} H., {Stern} D.,
  {Dey} A., 1997, AJ, 114, 2232

\bibitem[\protect\citeauthoryear{{Steidel} \& {Hamilton}}{{Steidel} \&
  {Hamilton}}{1993}]{SH93}
{Steidel} C.~C.,  {Hamilton} D., 1993, AJ, 105, 2017

\bibitem[\protect\citeauthoryear{{Stevenson}, {Shanks}, \& {Fong}}{{Stevenson}
  et~al.}{1986}]{SSF86}
{Stevenson} P.~R.~F., {Shanks} T.,  {Fong} R., 1986, in Spectral Evolution of
  Galaxies, p. 439

\bibitem[\protect\citeauthoryear{{Strecker}, {Erickson}, \&
  {Witteborn}}{{Strecker} et~al.}{1979}]{SEW79}
{Strecker} D.~W., {Erickson} E.~F.,  {Witteborn} F.~C., 1979, ApJS, 41, 501

\bibitem[\protect\citeauthoryear{{Teplitz}, {McLean}, \& {Malkan}}{{Teplitz}
  et~al.}{1999}]{Teplitz99}
{Teplitz} H.~I., {McLean} I.~S.,  {Malkan} M.~A., 1999, ApJ, 520, 469

\bibitem[\protect\citeauthoryear{{Thompson} et~al.}{{Thompson}
  et~al.}{1999}]{Thompsonetal99}
{Thompson} D. et~al., 1999, ApJ, 523, 100

\bibitem[\protect\citeauthoryear{{Tyson}}{{Tyson}}{1988}]{Tyson88}
{Tyson} J.~A., 1988, AJ, 96, 1

\bibitem[\protect\citeauthoryear{{Veron-Cetty} \& {Veron}}{{Veron-Cetty} \&
  {Veron}}{1996}]{VCV96}
{Veron-Cetty} M.-P.,  {Veron} P., 1996, "A Catalogue of quasars and active
  nuclei".
\newblock ESO Scientific Report, Garching: European Southern Observatory (ESO),
  7th ed.

\bibitem[\protect\citeauthoryear{{Wainscoat} \& {Cowie}}{{Wainscoat} \&
  {Cowie}}{1992}]{WC92}
{Wainscoat} R.~J.,  {Cowie} L.~L., 1992, AJ, 103, 332

\bibitem[\protect\citeauthoryear{{Williams} et~al.}{{Williams}
  et~al.}{2000}]{HDF00}
{Williams} R.~E. et~al., 2000, AJ, 120, 2735

\bibitem[\protect\citeauthoryear{{Williams} et~al.}{{Williams}
  et~al.}{1996}]{HDF96}
{Williams} R.~E. et~al., 1996, AJ, 112, 1335

\bibitem[\protect\citeauthoryear{{Yee} \& {Green}}{{Yee} \&
  {Green}}{1987}]{YG87}
{Yee} H.~K.~C.,  {Green} R.~F., 1987, ApJ, 319, 28

\end{thebibliography}

\label{lastpage}

\end{document}